\documentclass[12pt]{iopart}
\usepackage{iopams}

\newlength{\TZ}
\newcommand{\BEQ}{\begin{equation}}     
\newcommand{\BEA}{\begin{eqnarray}}
\newcommand{\EEQ}{\end{equation}}       
\newcommand{\EEA}{\end{eqnarray}}
\newcommand{\vph}{\varphi}              
\newcommand{\D}{{\rm d}}                
\newcommand{\wit}[1]{\widetilde{#1}}    

\renewcommand{\vec}[1]{\boldsymbol{#1}} 

\newcommand{\appsection}[2]{\setcounter{equation}{0}\setcounter{subsection}{0}
\section*{Appendix #1. #2}
\renewcommand{\theequation}{#1\arabic{equation}}
              \renewcommand{\thesection}{#1} }

\catcode`\@=11
\def\numberbysection{\@addtoreset{equation}{section}
        \def\theequation{\thesection.\arabic{equation}}}

\begin{document}

\title[LSI: bosonic contact and pair-contact processes]{Local 
scale-invariances in the bosonic contact and pair-contact processes}

\author{Florian Baumann$^{a,b}$, Stoimen Stoimenov$^{b,c}$ and 
Malte Henkel$^{b,d}$} 
\address{$^a$Institut f\"ur Theoretische Physik I, 
Universit\"at Erlangen-N\"urnberg, \\
Staudtstra{\ss}e 7B3, D -- 91058 Erlangen, Germany}
\address{$^b$Laboratoire de Physique des 
Mat\'eriaux,\footnote{Laboratoire associ\'e au CNRS UMR 7556} 
Universit\'e Henri Poincar\'e Nancy I, \\ 
B.P. 239, F -- 54506 Vand{\oe}uvre l\`es Nancy Cedex, France}
\address{$^c$Institute of Nuclear Research and Nuclear Energy,
Bulgarian Academy of Sciences, 1784 Sofia, Bulgaria}
\address{$^d$Isaac Newton Institute of Mathematical Sciences, 
20 Clarkson Road,\\ Cambridge CB3 0EH, England\footnote{Address until the 
$30^{\rm th}$ of April 2005}}

\begin{abstract}
Local scale-invariance for ageing systems without detailed balance is tested
through studying the dynamical symmetries of the critical bosonic contact process and the critical bosonic pair-contact process. 
Their field-theoretical
actions can be split into a Schr\"odinger-invariant term and a pure noise term. 
It is shown that the two-time response and correlation 
functions are reducible to certain multipoint response functions which depend
only on the Schr\"odinger-invariant part of the action. For the bosonic
contact process, the representation of the Schr\"odinger group can be 
derived from the free diffusion equation, whereas for the bosonic pair-contact 
process, a new representation of the Schr\"odinger group related to a non-linear Schr\"odinger equation with dimensionful couplings 
is constructed. The resulting
predictions of local scale-invariance 
for the two-time responses and correlators are completely
consistent with the exactly-known results in both models. 
\end{abstract}

\submitto{\JPA}
\maketitle

\setcounter{footnote}{0}

\section{Introduction}

The concept of scale-invariance is central to the modern understanding of
critical phenomena in and out of equilibrium. Its exploitation through the
renormalization group has in particular led to the recognition of universal
critical exponents and scaling functions which describe the behaviour of
physical observables, see e.g. \cite{Card96} and references therein. 
Here we are concerned with the slow dynamics of systems
brought rapidly to their critical point and/or into a phase with more than one
thermodynamically stable state. Such a kind of behaviour is typical for glassy
systems but also occurs in simple magnets with a purely relaxational dynamics 
which were quenched from a disordered state to a final temperature $T\leq T_c$, 
where $T_c>0$ is the critical temperature. 
For the latter, it is now understood that the dynamics is governed by a 
single time-dependent length-scale 
$L=L(t)\sim t^{1/z}$ for $t$ sufficiently large and where $z$ is 
the dynamical exponent \cite{Bray94a}. As an example, 
consider simple magnets relaxing towards equilibrium. 
For phase-ordering kinetics ($T<T_c$), the Bray-Rutenberg theory 
shows that dynamical scaling together with the assumption of a 
Porod law for the time-dependent structure factor predicts the value of 
$z$ \cite{Bray94}; whereas for $T=T_c$ the value of $z$ is computed from 
critical (equilibrium) dynamics. More recently, it has been understood that 
the study to two-time observables provides further and deeper insight, in
particular the ageing behaviour is made explicit through the breaking of 
time-translation invariance. The challenge is now to find the values of the 
associated non-equilibrium (ageing) exponents and also the form of the scaling 
functions, see below for the precise definitions. 

A common way to study this problem is through a Langevin equation which should
describe the dynamics of a coarse-grained order-parameter. This may be turned
into a field-theory and renormalization-group methods then allow to 
extract values of these exponents, in quite good but not perfect agreement 
with the results of direct numerical simulations \cite{Maze04,Cala05}. 
On the other hand, the resulting
predictions for the scaling functions appear to be far from the numerical
results, see \cite{Brow97,Henk04}. 
An alternative approach seeks to extend
dynamical scaling to a larger group of `local' scale-transformations
\cite{Henk02}, see \cite{Henk05} for a recent review. 
In the framework of a de Dominicis-Janssen type theory \cite{deDo78,Jans92}, the effective action $S=S[\phi,\wit{\phi}]$ is given in term of the 
order-parameter field $\phi$ and its
associated response field $\wit{\phi}$. Furthermore, for systems in contact
with a thermal bath such that detailed balance holds one always has the
decomposition $S[\phi,\wit{\phi}]=S_0[\phi,\wit{\phi}]+S_b[\wit{\phi}]$ into
a `deterministic' part $S_0$ which can be derived  from the Langevin
equation when all noise  terms are dropped and the `noise' term 
$S_b[\wit{\phi}]$ which depends only on the response function \cite{Pico04}. 
The form of response functions can then be found from the requirement of 
covariance under the group of local scale-transformations. 
As we shall explain below, 
correlation functions can be reduced to certain integrals of higher, 
multipoint response functions \cite{Pico04}. 
This approach yields the form of the scaling functions 
whereas the exponents are treated as parameters whose values have to be 
supplied\footnote{This is close in spirit to the treatment of {\em equilibrium}
phase transitions through conformal invariance, which fixes the form of the
$n$-point correlators in terms of the scaling dimensions of the scaling 
fields \cite{Card96}. Furthermore, those exponents can be determined exactly in 
$2D$ from symmetry considerations (i.e. representation theory of the Virasoro 
algebra) alone since the conformal symmetry is infinite-dimensional in that 
case.} and reproduce perfectly the
results of both analytical and simulational studies of many common spin systems
undergoing phase-odering kinetics where $z=2$.\footnote{For non-equilibrium 
critical dynamics where $z\ne 2$ in general one also has a good match with 
numerical data for the response functions in direct space but systematic 
differences may appear in momentum-space calculations, e.g. in the $2D$ Ising 
model for $t/s\lesssim 10$ \cite{Plei05}.} 

It is an established fact that the basic Langevin equation for the
order-parameter does not admit any symmetries beyond dilatations and 
(space-)translations, see \cite{Cham05} for a recent discussion. 
However, it has been shown that at least for simple 
magnets it is enough to concentrate on the dynamical symmetries 
of the {\em deterministic} part of
the Langevin equation only, as given by the action $S_0$. 
In particular, Schr\"odinger-invariance of that
deterministic part is sufficient to be able to derive the two-time
correlations $C(t,s)$ and two-time response functions $R(t,s)$ explicitly 
\cite{Pico04}. There is an exact agreement for systems such as the spherical 
model, the XY model in spin-wave approximation or the voter model which are all
described by a linear Langevin equation. Good agreement with simulations 
of Ising, Potts and XY models models was found as well 
\cite{Henk04,Abri04,Lore05}. 

In this paper, we extend the treatment of local scale-invariance to ageing systems with a dynamical exponent $z=2$ but without detailed balance. 
Working with a de Dominicis-Janssen type theory, we find again a decomposition 
$S[\phi,\wit{\phi}]=S_0[\phi,\wit{\phi}]+S_b[\phi,\wit{\phi}]$ into a `deterministic', Schr\"odinger-invariant term $S_0$ 
and `noise' terms, each of which contains at least one response field more
than order-parameter fields (explicit expressions will be given in sections~2
and~3). Then the Bargman superselection rules which
follow from the Galilei-invariance of $S_0$ are enough to establish that
again the two-time response function is noise-independent and the two-time
correlation function can be reduced to a finite sum of response functions
the form of whom is strongly constrained again by the requirement of their 
Schr\"odinger-covariance. 
These developments provide further evidence for {\em a hidden non-trivial local
scale-invariance in ageing systems} which manifests itself directly in the 
`deterministic' part (see \cite{Stoi05} for the construction of 
Schr\"odinger-invariant semi-linear kinetic equations) but which strongly
constrains the full noisy correlations and reponses.

We test the present framework of local scale-invariance in two
exactly solvable systems with a non-linear coarse-grained Langevin equation. 
A convenient set of models with a non-trivial ageing behaviour is furnished 
by the bosonic contact \cite{Houc02} and pair-contact processes \cite{Paes04}, both at criticality. These systems are defined as follows. Consider
a set of particles of a single species $A$ which move on the sites of a
hypercubic lattice in $d$ dimensions. On any site one may have an arbitrary
(non-negative) number of particles.\footnote{This property distinguishes the
models at hand from the conventional (`fermionic') contact and pair-contact 
processes whose critical behaviour is completely different.} 
Single particles may hop to a nearest-neighbour site with unit rate and in 
addition, the following single-site creation and annihilation processes 
are admitted 
\BEQ \label{gl:rates}
m A \stackrel{\mu}{\longrightarrow} (m+1) A \;\; , \;\; 
p A \stackrel{\lambda}{\longrightarrow} (p-\ell) A \;\; ; \;\;
\mbox{\rm with rates $\mu$ and $\lambda$} 
\EEQ
where $\ell$ is a positive integer such that $|\ell| \leq p$. 
We are interested in the following special cases:
\begin{enumerate}
\item {\em critical bosonic contact process:} $p=m=1$. Here only
$\ell = 1$ is possible. Furthermore
the creation and annihilation rates are set equal $\mu =\lambda$.
\item {\em critical bosonic pair-contact process:} $p=m=2$. We fix
$\ell=2$, set $2 \lambda = \mu$ and define the control parameter
\footnote{If instead we would treat a coagulation process $2A\to A$, 
where $\ell=1$, the results presented in the text are recovered by setting
$\lambda = \mu$ and $\alpha = {\mu}/{D}$.}
\BEQ
  \alpha := \frac{3 \mu}{2 D}
\EEQ
\end{enumerate}
The dynamics is described in terms of a master equation which may be written
in a hamiltonian form $\partial_t |P(t) \rangle = - H |P(t)
\rangle$ where $|P(t) \rangle$ is the time-dependent state
vector and the hamiltonian $H$ can be expressed in terms of
creation and annihilation operators $a(\vec{x},t)^\dag$ and
$a(\vec{x},t)$ \cite{Doi76,Schu00,Taeu05}. It is well-known that these
models are critical in the sense that their relaxation towards the steady-state
is algebraically slow \cite{Houc02,Paes04,Baum05}. In particular, the 
local particle-density is $\rho(\vec{x},t) := \langle a(\vec{x},t) \rangle$. 
Its spatial average remains constant in time
\BEQ
\int\!\D\vec{x}\, \rho(\vec{x},t) = 
\int\!\D\vec{x}\, \langle a(\vec{x},t) \rangle = \rho_0 
\EEQ
where $\rho_0$ is the initial mean particle-density. We are interested in the 
two-time connected correlation function
\BEQ
\label{def:correlator}
G(\vec{r};t,s) := \langle a(\vec{x},t) a(\vec{x}+\vec{r},s)\rangle - \rho_0^2
\EEQ
and take an uncorrelated initial state, hence $G(\vec{r};0,0)=0$. The linear 
two-time response function is found by adding a particle-creation term 
$\sum_{\vec{x}} h(\vec{x},t)\left(a^\dag(\vec{x},t)-1\right)$ 
to the quantum hamiltonian $H$ and taking the functional derivative
\BEQ
R(\vec{r};t,s) := 
\left.\frac{\delta \langle a(\vec{r}+\vec{x},t)\rangle}{\delta h(\vec{x},s)}
\right|_{h=0}
\EEQ
\begin{table}[t]
  \[
  \begin{array}{||c|c||c|c||}  \hline \hline
    & & \multicolumn{2}{c||}{\mbox{bosonic pair-contact process}} \\
         \cline{3-4}  
    & \raisebox{1.5ex}[-1.5ex]{ \mbox{bosonic contact process}}
    & \alpha < \alpha_c  & \alpha = \alpha_c  \\ \hline
    \hline
    a  & \frac{d}{2}-1 & \frac{d}{2}-1 & \frac{d}{2}-1  \\
    \hline
    b & \frac{d}{2}-1& \frac{d}{2}-1  & \begin{array}{ccl}
      0 & \mbox{if} & 2 < d < 4  \\
      \frac{d}{2}-2 & \mbox{if} & d > 4 
    \end{array}
      \\
    \hline
     \hline
  \end{array}
  \]
\caption[AgeingTab1]{Ageing exponents of the critical bosonic contact 
and pair-contact processes in the different regimes. The results for the 
bosonic contact process hold for an arbitrary dimension $d$, but for the
bosonic pair-contact process they only apply if $d>2$, since
$\alpha_c=0$ for $d\leq 2$.}
\label{tb:results_exp}
\end{table}
We have previously analyzed these quantities in the scaling limit
where both $t,s$ as well as $t-s$ become large with respect to some
microscopic reference time. The results are as follows \cite{Baum05}: 
consider the autocorrelation and autoresponse functions, which satisfy the
scaling forms 
\BEA
\label{gl:G}
G(t,s) &:=& G(\vec{0};t,s) = s^{-b} f_G(t/s) \\
\label{gl:R}
R(t,s) &:=& R(\vec{0};t,s) = s^{-1-a} f_R(t/s)
\EEA 
where the values of the exponents $a$ and $b$ are listed in 
table~\ref{tb:results_exp}. Here the critical value $\alpha_c$ for the
pair-contact process is explicitly given by \cite{Paes04}
\BEQ
\frac{1}{\alpha_c} = 2 \int_0^{\infty} \!\D u\, \left( e^{-4u} I_0(4u)\right)^d
\EEQ
where $I_0$ is a modified Bessel function. The dynamical behaviour of
the contact process is independent of $\alpha$. For the critical bosonic 
pair-contact process,
there is a clustering transition between a spatially homogeneous state
for $\alpha<\alpha_c$ and a highly inhomogeneous one for
$\alpha>\alpha_c$ where dynamical scaling does not hold. These
two transitions are separated by a multicritical point at
$\alpha=\alpha_c$. 
Since our models do not satisfy detailed balance, there is no reason why
the exponents $a$ and $b$ should co\"{\i}ncide and our result $a\ne b$ for
the bosonic pair-contact process is perfectly natural. 

While the scaling function $f_R(y)=(y-1)^{-d/2}$ has a very simple form,
the autocorrelator scaling function has an integral representation 
\BEQ
\label{gl:result_int}
\hspace{-1.0truecm}
f_G(y) = \mathcal{G}_0 \int_0^1 \!\D\theta\, 
\theta^{a-b} (y+1-2 \theta)^{-{d}/{2}}
\EEQ
where the values for $a$ and $b$ are given in 
table~\ref{tb:results_exp} and $\mathcal{G}_0$ is a known
normalization constant. The explicit scaling functions are listed up to
normalization in table~\ref{tb:results_fun} \cite{Baum05}. In this 
paper, we shall study to what extent their form can be understood from local 
scale-invariance. 

\begin{table} 
  \[
  \hspace*{-2.5cm}
  \begin{array}{||c|c|c||c|c||}  \hline \hline
    \multicolumn{3}{||c||}{} &  f_R(y) & f_G(y) \\
    \hline
    \hline
    \multicolumn{3}{||c||}{\mbox{contact process}} &
    (y-1)^{-\frac{d}{2}}  & (y-1)^{-\frac{d}{2}+1} -
    (y+1)^{-\frac{d}{2}+1} \\ \hline \hline
    \mbox{pair} & \alpha < \alpha_c & d > 2 & (y-1)^{-\frac{d}{2}} &
    (y-1)^{-\frac{d}{2}+1} - (y+1)^{-\frac{d}{2}+1}  \\ \cline{2-5}
    \mbox{contact} &  & 2 < d < 4 & (y-1)^{-\frac{d}{2}} & 
    (y+1)^{-\frac{d}{2}} {_2F_1}
    \left(\frac{d}{2},\frac{d}{2};\frac{d}{2}+1;\frac{2}{y+1}\right)\\
    \cline{3-5}
    \mbox{process} & \raisebox{1.6ex}[-1.6ex]{$\alpha =
    \alpha_c$} & d > 4 & (y-1)^{-\frac{d}{2}} &  (y+1)^{-\frac{d}{2}+2}
    - (y-1)^{-\frac{d}{2}+2} + ( d-4 ) (y-1)^{-\frac{d}{2}+1} 
\\ \hline \hline
  \end{array}
  \]
\caption{Scaling functions (up to normalization) of the autoresponse and 
autocorrelation of the critical bosonic contact and bosonic pair-contact
processes.}
\label{tb:results_fun}
\end{table}

This paper is organized as follows. In section~2 we 
treat the bosonic contact process in its field-theoretical
formulation. The action is split into a Schr\"odinger-invariant term $S_0$
and a noise term $S_b$ and we show how the response and correlation functions
can be exactly reduced to certain noiseless three- and four-point
response functions. In this reduction the Bargman superselection rules
which follow from the Schr\"odinger-invariance of $S_0$ play a central r\^ole. These tools allow us to predict the reponse- and correlation
functions which will be compared to the exact results of
table~\ref{tb:results_fun}. In section~3 the same programme is carried
out for the bosonic pair-contact process but as we shall see, the 
Schr\"odinger-invariant term $S_0$ of its action is now related to a
{\em non-}linear Schr\"odinger equation. The treatment of this requires
an extension of the usual representation of the Schr\"odinger Lie-algebra
which now includes a dimensionful coupling constant. 
The construction is carried out in appendix~A. The required $n$-point 
correlation functions coming from this new representation are derived in 
appendices~B and C. 
Finally, in section~4 we conclude.

\section{The contact process}
\subsection{Field-theoretical description}

The master equation which describes the critical bosonic contact process 
as defined in section~1 can be turned into a field-theory in a 
standard fashion
through an operator formalism which uses a particle annihilation operator
$a(\vec{r},t)$ and its conjugate $a^{\dag}(\vec{r},t)$, see
for instance \cite{Doi76,Taeu05} for detailed discussion of the
technique. Since we shall be
interested in the connected correlator, we consider the shifted field and 
furthermore introduce the shifted response field
\BEA
\phi(\vec{r},t) &:=& a(\vec{r},t) - \rho_0 \nonumber \\
\wit{\phi}(\vec{r},t) &:=& \bar{a}(\vec{r},t) 
= a^{\dag}(\vec{r},t) - 1
\label{gl:phiphit}
\EEA
such that $\langle \phi(\vec{r},t) \rangle = 0$ (our notation implies a mapping
between operators and quantum fields, using the known equivalence between 
the operator formalism and the path-integral
formulation \cite{Droz94,Taeu05}). As we shall see, these
fields $\phi$ and $\wit{\phi}$ will become the natural quasiprimary fields
from the point of view of local scale-invariance. 
We remark that the response function is not affected by
this shift, since 
\BEQ
R(\vec{r},\vec{r}';t,s) = \frac{\delta \langle a(\vec{r},t)
\rangle}{\delta h(\vec{r}',s)} = 
\frac{\delta \langle \phi(\vec{r},t)
\rangle}{\delta h(\vec{r}',s)} 
\EEQ
Then the field-theory action reads, where $\mu$ is the
reaction rate \cite{Howa97}
\BEA
\label{gl:cpd:action2}
S[\phi,\wit{\phi}] 
&=& \int \!\D \vec{R} \int \!\D u \: \left[ \wit{\phi}
(2\mathcal{M}\partial_u - \nabla^2)\phi - \mu
{\wit{\phi}}^2 (\phi+\rho_0) \right]
\nonumber \\
&=& S_0[\phi,\wit{\phi}] +  S_b[\phi,\wit{\phi}] 
\EEA
To keep expressions shorter, we have supressed the arguments of $\phi(\vec{R},u)$ and $\tilde{\phi}(\vec{R},u)$ under the
integrals and we shall also do so often in what follows, if no ambiguity arises.
The diffusion constant $D$ is related to the `mass' $\mathcal{M}$ through
$D=(2\mathcal{M})^{-1}$. We have decomposed the action as 
follows: 
\BEQ
S_0[\phi,\wit{\phi}] := \int \!\D \vec{R} \int \!\D u\:
\left[ \wit{\phi} (2\mathcal{M}\partial_u - \nabla^2)\phi \right]
\EEQ
describes the deterministic,\footnote{This terminology is used since the
equation of motion of $\phi$ following from $S_0$ is a partial differential equation and
not a stochastic Langevin equation.} noiseless part whereas the noise is described by 
\BEQ
S_b[\phi,\wit{\phi}] := - \mu \int \!\D \vec{R} \int \!\D u\: 
\left[{\wit{\phi}}^2 (\phi+\rho_0) \right].
\EEQ
quite analogously to what happens in the kinetics of simple magnets, see
\cite{Pico04} for details. 

In principle, an initial correlator $G(\vec{r};0,0)$
could be assumed and will lead to a further contribution $S_{\rm ini}$ to the
action. For critical systems, one usually employs a term of the form
$S_{\rm ini,st}=-\frac{\tau_0}{2} \int \!\D \vec{R}\, (\phi(\vec{R},0) - 
\langle \phi(\vec{R},0) \rangle)^2$, see e.g. \cite{Jans92,Cala05} but this would have for us the disadvantage that it explicitly breaks Galilei-invariance. 
We shall rather make use of the Galilei-invariance of the 
noiseless action $S_0[\phi,\wit{\phi}]$ and use as an 
initial term \cite{Maze04,Pico04}
\BEQ
S_{\rm ini}[\wit{\phi}] = -\frac{1}{2} \int \!\D \vec{R} \D \vec{R'}\;
\wit{\phi}(\vec{R},0) G(\vec{R}-\vec{R'};0,0)
\wit{\phi}(\vec{R}',0).
\EEQ
Because of the initial condition $G(\vec{R};0,0)=0$, however, 
$S_{\rm ini}[\wit{\phi}]=0$ and we shall not need to consider it any further. 

{}From the action (\ref{gl:cpd:action2}), $n$-point functions
can then be computed as usual 
\BEQ
\hspace{-2.0truecm}
\langle \phi_1(\vec{r}_1,t_1) \ldots \phi_n(\vec{r}_n,t_n)
\rangle = \int \mathcal{D} \phi \mathcal{D} \wit{\phi} \; 
\phi_1(\vec{r}_1,t_1) \ldots
\phi_n(\vec{r}_n,t_n)  \exp\left(
-S[\phi,\wit{\phi}] \right)
\EEQ
which through the decomposition (\ref{gl:cpd:action2}) can be written as an 
average of the noiseless theory 
\BEQ
\hspace{-1.0truecm} 
\langle \phi_1(\vec{r}_1,t_1) \ldots \phi_n(\vec{r}_n,t_n)
\rangle = \left\langle \phi_1(\vec{r}_1,t_1) \ldots
\phi_n(\vec{r}_n,t_n) \exp
\left(-S_b[\phi,\wit{\phi}]\right) \right\rangle_0
\EEQ
where $\langle \ldots \rangle_0$ denotes the
expectation value with respect to the noiseless theory.

\subsection{Symmetries of the noiseless theory}
\label{symmetries}

In what follows, we shall need some symmetry properties of the noiseless
part described by the action $S_0[\phi,\wit{\phi}]$ 
which we now briefly recall. The noiseless equation of motion for the 
field $\phi$ is a free diffusion-equation 
$2\mathcal{M}\partial_t \phi(\vec{x},t) = \nabla^2 \phi(\vec{x},t)$. 
Its dynamical symmetry group is the well-known 
Schr{\"o}dinger-group {\sl Sch}$(d)$ \cite{Lie1882,Nied72} which acts on 
space-time coordinates $(\vec{r},t)$ as 
$(\vec{r},t)\mapsto (\vec{r}',t') = g(\vec{r},t)$ where 
\BEQ
t \longrightarrow t' = \frac{\alpha t + \beta}{\gamma t +
\delta},\;\;\; \vec{r} \longrightarrow \vec{r}' =
\frac{\mathcal{R} \vec{r} + \vec{v} t + \vec{a}}{\gamma t +
\delta}; \;\;\; \alpha \delta - \beta \gamma = 1
\EEQ
and where $\mathcal{R}$ is a rotation matrix. Solutions $\phi$ of
the free diffusion equation are carried to other
solutions of the same equation and $\phi$ transforms as
\BEQ
\label{gl:field_transf}
\phi(\vec{r},t) \longrightarrow (T_g \phi)(\vec{r},t) =
f_{g}[g^{-1}(\vec{r},t)] \phi[g^{-1}(\vec{r},t)]
\EEQ
where the companion function $f_{g}$ is known explicitely
and contains the so-called `mass' ${\cal M}=(2D)^{-1}$ \cite{Nied72,Nied74}. 
We list the generators of the Lie algebra 
$\mathfrak{sch}_1 = {\rm Lie}(\mbox{\sl Sch}(1))$ in one spatial 
dimension \cite{Henk94}
\BEA
X_{-1} & = & -\partial_t  \nonumber \\
X_0 & = & -t \partial_t - \frac{1}{2} {r}
\partial_{{r}} - \frac{x}{2} \nonumber \\
X_1 & = & - t^2 \partial_t - t {r} \partial_{{r}} -
x t - \frac{\mathcal{M}}{2} {r}^2 \nonumber \\
Y_{-\frac{1}{2}} & = & - \partial_{{r}} 
\label{gl:sch1gen} \\
Y_{\frac{1}{2}}  & = & - t \partial_{{r}} - \mathcal{M} {r} \nonumber \\
M_0 & = & - \mathcal{M} \nonumber
\EEA
Fields transforming under {\sl Sch}$(d)$ are characterized by a scaling 
dimension and a mass. We list in table~\ref{tab:champs} some fields which we 
shall use below. 
\begin{table}
\begin{center}
\begin{tabular}{|c|cr|} \hline 
field & scaling dimension & mass \\ \hline 
$\phi$       & $x$ & $\cal M$ \\
$\wit{\phi}$ & $\wit{x}$ & $-{\cal M}$ \\
${\wit{\phi}}^2$ & $\wit{x}_2$ & $-2{\cal M}$ \\
$\Upsilon := {\wit{\phi}}^2 \phi$ & $x_{\Upsilon}$ & $-{\cal M}$ \\ 
$\Sigma :=  {\wit{\phi}}^3 \phi$ & $x_{\Sigma}$ & $-2 {\cal M}$  \\ 
$\Gamma:=  {\wit{\phi}}^3 \phi^2$ & $x_{\Gamma}$ & $-{\cal M}$  \\ \hline
\end{tabular} \end{center}
\caption{Scaling dimensions and masses of some composite fields.
\label{tab:champs}}
\end{table}
We remark that for free fields one has
\BEQ
\wit{x}_2 = 2 \wit{x} \;\; , \;\;
x_{\Upsilon} = 2 \wit{x}+x\;\; , \;\;
x_{\Sigma} = 3\wit{x} + x \;\; , \;\;
x_{\Gamma} = 3\wit{x} + 2 x
\EEQ
but these relations need no longer hold for interacting fields. On the other
hand, from the Bargman superselection rules (see \cite{Barg54} and below) we
expect that the masses of the composite fields as given in
table~\ref{tab:champs} should remain valid for interacting fields as well. 

Throughout this paper, we shall make the important assumption that the
fields $\phi$ and $\wit{\phi}$ transform covariantly according to
(\ref{gl:field_transf}) under the Schr\"odinger group. By analogy with
conformal invariance, such fields are called {\em quasiprimary} \cite{Henk02}. 
For quasiprimary fields the so-called Bargman superselection rules \cite{Barg54}
holds true which state that
\BEQ
\label{gl:bargman1}
\langle \: \underbrace{\phi \ldots \phi}_{n}\:
\underbrace{\wit{\phi}\ldots \wit{\phi}}_{m} \:\rangle_0 = 0  
\;\; \mbox{\rm ~~unless $n=m$}
\EEQ
We recall the proof of these in appendix~B. 
Before we consider the consequences of (\ref{gl:bargman1}), 
we recall the well-known result on the form of noise-less $n$-point 
functions in ageing systems.  

Since in ageing phenomena, time-translation invariance is broken, we must 
consider the subalgebra $\mathfrak{age}_1\subset\mathfrak{sch}_1$ obtained by 
leaving out the generator of time-translations $X_{-1}$ \cite{Henk03}. Then the
$n$-point function of quasiprimary fields $\phi_i$, $i=1,\ldots n$ has to 
satisfy the covariance conditons \cite{Henk94,Henk02}
\BEA
\label{gl:cpd:infinitesimal1}
\left(\sum_{i=1}^n X_k^{(i)}\right)
\langle \vph_1(\vec{r}_1,t_1) \ldots \vph_n(\vec{r}_n,t_n) \rangle_0 &=& 0 
\;\; ; \;\;  k \in \{0,1\} \\ 
\label{gl:cpd:infinitesimal2}
\left(\sum_{i=1}^n Y_m^{(i)}\right)
\langle \vph_1(\vec{r}_1,t_1) \ldots \vph_n(\vec{r}_n,t_n)
\rangle_0  &=& 0 \;\; ; \;\;
m \in \left\{-\frac{1}{2},\frac{1}{2} \right\}
\EEA
where $\vph_i$ stands either for a quasiprimary field $\phi_i$ or 
a quasiprimary response field $\wit{\phi}_i$. The $\vph_i$ are characterized by their scaling dimension $x_i$ and their mass $\mathcal{M}_i$. 
The generators $X_k$ are then the extension of
(\ref{gl:sch1gen}) to $n$-body operators and the superscript $(i)$ refers to 
$\vph_i$. The $n$-point function is zero
unless the sum of all masses vanishes 
\BEQ
\sum_{i=1}^n {\cal M}_i = 0
\EEQ
which reproduces the Bargman superselection rule (\ref{gl:bargman1}). 
It is well-known \cite{Henk94,Henk02} that the noiseless two-point function 
$R_0(\vec{r},\vec{r}';t,s)=\langle\vph_1(\vec{r},t)\vph_2(\vec{r}´,s)\rangle_0$ 
is completely determined by the equations (\ref{gl:cpd:infinitesimal1}) and
(\ref{gl:cpd:infinitesimal2}) up to a normalization constant.
\BEQ
\hspace{-1.0em}
R_0(\vec{r},\vec{r}';t,s) = R_0(t,s) \exp \left( -\frac{\mathcal{M}_1}{2}
\frac{(\vec{r}-\vec{r}')^2}{(t-s)} \right) \, 
\delta(\mathcal{M}_1+\mathcal{M}_2)
\EEQ
where the autoresponse function is given by 
\BEQ
R_0(t,s) = r_0
(t-s)^{-\frac{1}{2}(x_1 + x_2)} \left( \frac{t}{s} \right)^{-\frac{1}{2}(x_1 -
x_2)}
\EEQ
This reproduces the expected scaling form (\ref{gl:R}) together with the
scaling function $f_R(y)$ as given in table~\ref{tb:results_fun} if we identify
\BEQ \label{gl:x1x2a}
x = x_1 = x_2, \qquad \mbox{and} \qquad x = a + 1
\EEQ
For the critical bosonic contact process, we read off from 
table~\ref{tb:results_exp} that $a = \frac{d}{2}-1$. Hence one recovers 
$x =\frac{d}{2}$, as expected for a free-field theory.

\subsection{Reduction formul{\ae}}
\label{reduction}

We now show that the Bargman superselection rule (\ref{gl:bargman1}) 
implies a reduction
of the $n$-point function of the full theory to certain correlators of the
noiseless theory, which is described by $S_0$ only. This can be done 
generalizing the arguments of \cite{Pico04}. 

First, for the computation of the response function, we add the
term $\int \!\D \vec{R} \int \D u\,  \wit{\phi}(\vec{R},u)
h(\vec{R},u)$ to the action. As usual the response function is
\BEA
\lefteqn{R(\vec{r},\vec{r}';t,s) = \left\langle \phi(\vec{r},t)
\wit{\phi}(\vec{r}',s) \right\rangle} \nonumber \\
&=& \left \langle \phi(\vec{r},t)
\wit{\phi}(\vec{r}',s) \exp \left( - \mu \int \!\D \vec{R}
\int \!\D u\: \wit{\phi}^2(\vec{R},u) 
(\phi(\vec{R},u) + \rho_0) \right) \right \rangle_0 
\nonumber \\
&=& \left\langle \phi(\vec{r},t) \wit{\phi}(\vec{r}',s) \right\rangle_0
= R_0(\vec{r},\vec{r'};t,s)
\label{gl:Rsansbruit}
\EEA
where we expanded the exponential and applied the Bargman
superselection rule. Indeed, the two-time response is just given by the
response of the (gaussian) noise-less theory. We have therefore
reproduced the exact result of table~\ref{tb:results_fun} 
for the response function of the critical bosonic contact process.

Second, we have for the correlator
\BEA
 G(\vec{r},\vec{r}',t,s) &= &\left\langle \phi(\vec{r},t)
\phi(\vec{r}'s) \exp \left( -\mu \int \!\D \vec{R} \int \!\D u\:
\wit{\phi}^2(\vec{R},u) \phi(\vec{R},u) \right) \right.
\nonumber \\ 
& & \left. ~~ \times \exp \left( -\mu \rho_0 \int \!\D \vec{R} \int \!\D u\:
\wit{\phi}^2(\vec{R},u) \right) \right\rangle_0
\EEA
Expanding both exponentials
\BEA
& &\hspace{-2.0truecm}
\exp \left( -\mu \int \!\D \vec{R} \int \!\D u\:
\wit{\phi}^2(\vec{R},u) \phi(\vec{R},u) \right) =
\sum_{n=0}^\infty \frac{(-\mu)^n}{n!} \left( \int \!\D \vec{R} \int \!\D u\:
\wit{\phi}^2(\vec{R},u) \phi(\vec{R},u) \right)^n
\nonumber \\
& & \hspace{-2.0truecm} \exp \left( -\mu \rho_0\int \!\D \vec{R} \int \!\D u\:
\wit{\phi}^2(\vec{R},u) \right) =
\sum_{m=0}^\infty \frac{(-\rho_0 \mu)^m}{m!} 
\left( \int \!\D \vec{R} \int \!\D u\:
\wit{\phi}^2(\vec{R},u) \right)^m \nonumber
\EEA
and using the Bargman superselection rule (\ref{gl:bargman1}), 
non-vanishing terms only arise if $2 n  + 2m = n + 2$ or else 
\BEQ
n + 2m = 2
\EEQ
This can only be satisfied for $n = 0$ and $m = 1$ or for $n
= 2$ and $m = 0$. The full noisy correlator hence is the sum of only two terms
\BEQ
G(\vec{r},\vec{r}';t,s) = G_1(\vec{r},\vec{r}';t,s) +
G_2(\vec{r},\vec{r}';t,s)
\EEQ
where the first contribution involves a three-point function
of the composite field $ \wit{\phi}^2$ of
scaling dimension $\tilde{x}_2$ (see table~\ref{tab:champs})
\BEQ
\label{gl:cont1}
G_1(\vec{r},\vec{r}';t,s) = - \mu \rho_0 \int \!\D \vec{R}
\int \!\D u\: \left\langle \phi(\vec{r},t) \phi(\vec{r}',s)
\wit{\phi}^2(\vec{R},u) \right\rangle_0
\EEQ
whereas the second contribution comes from a four-point
function and involves the composite field $\Upsilon$ 
(see table~\ref{tab:champs})
\BEQ
\hspace{-2.0truecm}
G_2(\vec{r},\vec{r}';t,s) = \frac{\mu^2}{2} \int \!\D
\vec{R} \D \vec{R}' \int \!\D u \D u'\: \left\langle \phi(\vec{r},t)
\phi(\vec{r}',s) \Upsilon(\vec{R},u) \Upsilon(\vec{R}',u') \right\rangle_0
\EEQ
We see that the connected correlator is determined by three- 
and four-point functions of the noiseless theory. We now use the
symmetries of that noise-less 
theory to determine the two-, three,- and four-point functions
as far as possible. 

\subsection{Correlator with noise}

We consider $G_1(\vec{r},\vec{r}',t,s)$ first. The
appropriate three-point function is given in appendix~B, 
equation (\ref{app_threepoint}): 
\BEA
\langle \phi(\vec{r},t) \phi(\vec{r}',s) \wit{\phi}^2
(\vec{R},u) \rangle_0 &=&
(t-s)^{x-\frac{1}{2}\tilde{x}_2} (t-u)^{-\frac{1}{2} \tilde{x}_2}
(s-u)^{-\frac{1}{2} \tilde{x}_2} \nonumber \\ & &
\hspace{-5.5truecm} \times \exp \left( -\frac{\mathcal{M}}{2}
\frac{(\vec{r}-\vec{R})^2}{t-u} - \frac{\mathcal{M}}{2}
\frac{(\vec{r}'-\vec{R})^2}{s - u} \right) \Psi_3 ( u_1,v_1)
\Theta(t-u) \Theta(s-u)
\EEA
with
\BEA
u_1 & = & \frac{u}{t}\cdot \frac{ [(s-u)(\vec{r}-\vec{R})
- (t-u)(\vec{r}' - \vec{R})]^2 }{(t-u) (s-u)^2} \nonumber \\
v_1 & = & \frac{u}{s}\cdot \frac{ [(s-u)(\vec{r}-\vec{R})
- (t-u)(\vec{r}' - \vec{R})]^2 }{(t-u)^2 (s-u)} 
\EEA
and an undetermined scaling function $\Psi_3$. The
$\Theta$-functions have been introduced by hand because of
causality but this could be justified through a more elaborate 
argument along the lines of \cite{Henk03}. 
Introduced into (\ref{gl:cont1}), this gives the general form for the
contribution $G_1(\vec{r},\vec{r}';t,s)$. We concentrate
here on the autocorrelator, i.e. $\vec{r} = \vec{r}'$ and find,
with $y=t/s$
\BEA
\lefteqn{G_1(t,s) = -\mu \rho_0\, s^{-x - \frac{1}{2}\tilde{x}_2 +
\frac{d}{2} + 1} \cdot (y-1)^{-(x - \frac{1}{2} \tilde{x}_2)}} 
\nonumber \\ 
& & \hspace{-0.5cm} \times \int_0^1 \!\D
\theta\, (y-\theta)^{-\frac{1}{2} \tilde{x}_2}
(1-\theta)^{-\frac{1}{2}
\tilde{x}_2} \int_{\mathbb{R}^d} \!\D \vec{R}\, \exp \left(
-\frac{\mathcal{M}}{2} \vec{R}^2
\frac{y+1-2\theta}{(y-\theta)(1-\theta)} \right)\nonumber
\\& & \hspace{-0.5cm} \times H \left(
\frac{\theta}{y}
\frac{\vec{R}^2 (y-1)^2}{(y-\theta)(1-\theta)^2}, \theta\,
\frac{\vec{R}^2 (y-1)^2}{(y-\theta)^2 (1-\theta)} \right)
\EEA
where $H$ is an undetermined scaling function. Very much in the same way, 
we find for $G_2(t,s)$ 
\BEA
& &\hspace{-2.0truecm}
G_2(t,s) =  \frac{\mu^2}{2}\, s^{-x-x_{\Upsilon}+d+2} \cdot 
(y-1)^{-(x-x_{\Upsilon})} \int_0^1 \!\D
\theta \int_0^1 \!\D \theta'\, (y-\theta)^{-\frac{1}{2}
x_{\Upsilon}} (1-\theta)^{-\frac{1}{2} x_{\Upsilon}}\nonumber \\ 
& & \hspace{-1.0truecm} \times (y-\theta')^{-\frac{1}{2}
x_{\Upsilon}} (1-\theta')^{-\frac{1}{2} x_{\Upsilon}} 
\int_{\mathbb{R}^{2d}} \!\D \vec{R} \D \vec{R}'\,
\exp\left( -\frac{\mathcal{M}}{2}
\frac{\vec{R}^2}{1-\theta} -\frac{\mathcal{M}}{2}
\frac{\vec{R}'^2}{1-\theta'} \right) \nonumber \\ & &
\hspace{-1.0truecm} \times \Psi_4 \left(
\tilde{u}_3(\vec{R},\theta,\vec{R}',\theta'),
\tilde{u}_4(\vec{R},\theta,\vec{R}',\theta'),
\tilde{v}_3(\vec{R},\theta,\vec{R}',\theta'),
\tilde{v}_4(\vec{R},\theta,\vec{R}',\theta') \right)
\EEA
where $\Psi_4$ is another undetermined function and the
functions $\tilde{u}_3,\tilde{u}_4,\tilde{v}_3,\tilde{v}_4$
can be worked out from the appropriate expressions
(\ref{app_fourpoint}) in the
appendix B by the replacements $\mathbf{r}_3 - \mathbf{r}_2
\rightarrow \mathbf{R}$, $\mathbf{r}_4 - \mathbf{r}_2
\rightarrow \mathbf{R}'$, $t_2 \rightarrow 1$, $t_1 \rightarrow
y$, $t_3 \rightarrow \theta$, $t_4 \rightarrow \theta'$
(remember that $\vec{r}_1 = \vec{r}_2$)

As we have a free-field theory for the critical bosonic contact process,
we expect from table~\ref{tb:results_exp} and eq.~(\ref{gl:x1x2a}) that 
$x=\wit{x}=d/2$ and hence 
the following scaling dimensions for the composite fields
\BEQ
\tilde{x}_2 = d, \qquad x_{\Upsilon} = \frac{3}{2} d
\EEQ
Consequently, the autocorrelator takes the general form
\BEQ
G(t,s) = s^{1-d/2} g_1(t/s) + s^{2-d} g_2(t/s)
\EEQ
For $d$ larger than the lower critical dimension $d_*=2$, the second term
merely furnishes a finite-time correction. On the other hand, for $d<d_*=2$,
it would be the dominant one and we can only achieve agreement with the
known exact result if we assume $\Psi_4=0$. In what follows, we shall discard
the scaling function $g_2$ and shall concentrate on showing that our expressions
for $g_1$ are compatible with the exact results given in 
table~\ref{tb:results_fun}. 

In order to do so, we choose the following special form for the 
function $\Psi_3$
\BEQ \label{gl:3:Xi}
\Psi_3(u_1,v_1) = \Xi \left(\frac{1}{u_1}-\frac{1}{v_1}\right)
\EEQ
where $\Xi$ remains an arbitrary function. 
Then we are back in the case already treated in \cite{Pico04}. We find
\BEA
G_1(t,s) &=& -\mu \rho_0 s^{\frac{d}{2}+1-x-\frac{1}{2}\tilde{x}_2}
(y-1)^{\frac{1}{2} \tilde{x}_2 - x - \frac{d}{2}} 
\nonumber \\
& & \times \int_0^1 \!\D \theta\,
[(y-\theta)(1-\theta)]^{\frac{d}{2}-\frac{1}{2} \tilde{x}_2}
\phi_1 \left(\frac{y+1-2 \theta}{y-1} \right)
\EEA
where the function $\phi_1$ is defined by 
\BEQ
\phi_1(w) = \int \!\D \vec{R}\, \exp\left(
-\frac{\mathcal{M}w}{2} \vec{R}^2 \right) \Xi(\vec{R}^2)
\EEQ
As in \cite{Pico04} we choose
\BEQ
\phi_1(w) = \phi_{0,c} w^{-1-a}. 
\EEQ
This form for $\phi_1(w)$ guarantees
that the three-point response function 
\newline \typeout{ *** hier ist ein Zeilenvorschub !! *** } 
$\langle\phi(\vec{r},t) \phi(\vec{r},s) \phi^2(\vec{r}',u) \rangle_0$ is nonsingular for $t = s$.  We have thus
\BEQ
G(t,s) = G_1(t,s) = s^{-b} f_G\left(\frac{t}{s}\right)
\EEQ
with
\BEA
f_G(y) &=& -\mu \rho_0 \phi_{0,c} \int_0^1 \!\D \theta\, (y + 1 -
2 \theta)^{-\frac{d}{2}}\nonumber \\ 
&=& \frac{2 \mu \rho_0 \phi_{0,c}}{d}\left( (y-1)^{-\frac{d}{2} +1 } -
(y+1)^{-\frac{d}{2}+1} \right)
\EEA
and we have reproduced the corresponding entry in table~\ref{tb:results_fun}
for the critical bosonic contact process.\footnote{We remark that for $2<d<4$,
the same form of the autocorrelation function is also found in the critical
voter-model \cite{Dorn98}.}  

\section{The pair-contact process}

\subsection{Field-theoretical description and reduction formula}

For the pair-contact process we have two different cases, namely the case 
$\alpha < \alpha_c$ and the case at criticality $\alpha = \alpha_c$.  
The following considerations apply to both cases and we
shall for the moment leave the value of $\alpha$ arbitrary and only
fix it at a later state.

The action for the pair-contact process on the critical line
is \cite[eq. (30)]{Howa97}
\BEA
\hspace{-1.0cm}
S[a,\bar{a}] &=& \int \D \vec{R} \int \D u \left[\bar{a}
(2\mathcal{M}\partial_t - \nabla^2)a - \alpha \bar{a}^2 a^2 
- \mu \bar{a}^3a^2 \right]
\EEA
As before, see eq.~(\ref{gl:phiphit}), we switch to the quasiprimary fields
$\phi(\vec{r},t) = a(\vec{r},t) - \rho_0$ and
$\wit{\phi}(\vec{r},t) = \bar{a}(\vec{r},t)$. 
Then the action becomes 
\BEA
S[\phi,\wit{\phi}] &=& \int \!\D \vec{R} \int \!\D u\: \left[
\wit{\phi}(2\mathcal{M}\partial_t - \nabla^2) \phi - \alpha
\wit{\phi}^2 \phi^2 -  \alpha \rho_0^2 
\wit{\phi}^2 - \right.  \nonumber \\ 
& & \hspace{2.0cm} - \left.2 \alpha \rho_0\wit{\phi}^2  \phi -
\mu \wit{\phi}^3 \phi^2 -  2 \mu \rho_0 \wit{\phi}^3 \phi -
\mu \rho_0^2 \wit{\phi}^3 \right] \nonumber \\
&=& S_0[\phi,\wit{\phi}] + S_b[\phi,\wit{\phi}]
\EEA
Also in this model, similarly to the treatment of section~2, 
a decomposition of the action into a
first term with a non-trivial dynamic symmetry and a remaining noise term
is sought such that the correlators and responses can be reexpressed
in terms of certain $n$-point functions which only depend on $S_0$. 
The first term reads 
\BEQ
\label{gl:sigma02}
S_0[\phi,\wit{\phi}] := \int \!\D \vec{r} \int \!\D t\:
\left[ \wit{\phi} (2\mathcal{M}\partial_t - \nabla^2)\phi - \alpha\wit{\phi}^2
\phi^2\right].
\EEQ
and we derive its Schr\"odinger-invariance in appendix~A. 
The remaining part is the noise-term which reads
\BEA
\label{gl:kappa2}
\hspace{-1.8cm}
S_b[\phi,\wit{\phi}] & = & \int \!\D \vec{R} \int \!\D u\:
\left[- \alpha \rho_0^2 \wit{\phi}^2 - 2 \alpha \rho_0\wit{\phi}^2  \phi
-\mu \wit{\phi}^3 \phi^2- 2 \mu \rho_0
\wit{\phi}^3 \phi - \rho_0^2 \wit{\phi}^3 \right]
\EEA
Also in this case the Bargman superselection rule (\ref{gl:bargman1}) holds
true. This means that we can proceed now in a very similar way as before.\footnote{This argument works provided each term in $S_b$
contains at least one response field $\wit{\phi}$ more than order-parameter
fields $\phi$.}
First we have to check which $n$-point functions contribute to the
response and correlation funtion. We rewrite
$\exp(-S_b[\phi,\wit{\phi}])$ as a product of five exponentials
and expand each factor. The indices of the sums are
denoted by $k_i$ for the $i$-th term in (\ref{gl:kappa2}),
for instance for the first term
\BEQ
\hspace{-2.0truecm}
\exp \left(- \int \!\D \vec{R} \int \!\D u\:
\alpha \rho_0^2 \wit{\phi}^2(\vec{R},u) \right) =
\sum_{k_1 = 0}^\infty \frac{1}{k_1 !} \left( -\int \!\D \vec{R} \int \!\D u\:
\alpha \rho_0^2 \wit{\phi}^2(\vec{R},u) \right)^{k_1}
\EEQ
For the response function again only the
first term of each sum contributes, that is 
\BEQ \label{gl:4:R}
R(\vec{r},\vec{r}';t,s) = R_0(\vec{r},\vec{r}';t,s)
\EEQ
is noise-independent. 
For the correlation function, we have the condition 
$2 k_1 + 2 k_2 + 3 k_3 + 3 k_4 + 3 k_5 = 2 + k_2 + 2 k_3 +k_4$
or simply
\BEQ
2 k_1 + k_2 + k_3 +2 k_4 + 3 k_5 = 2
\EEQ
which implies immediately that
\BEQ
k_5 = 0.
\EEQ
In table~\ref{tb:contributions} we list the five differenent contributions
to the correlation function. We denote also the form of the
composite field, its scaling dimension and whether it is a
three- or four-point function which contributes.
\begin{table}
\[
\begin{array}{||c|c|c|c|c|c|c|c||} \hline \hline
 \mbox{contr.} & k_1 & k_2 & k_3 & k_4 &
  \mbox{comp.field} & \mbox{scaling dim.} &
  \mbox{3-point/4-point} \\ \hline
 G_1(t,s)& 1 & 0 & 0 & 0 & \wit{\phi}^2     &  \tilde{x}_2 & \mbox{3-point} \\
 G_2(t,s)& 0 & 2 & 0 & 0 & \Upsilon   & x_{\Upsilon} & \mbox{4-point} \\
 G_3(t,s)& 0 & 0 & 2 & 0 & \Gamma & x_{\Gamma} & \mbox{4-point} \\
 G_4(t,s)& 0 & 0 & 0 & 1 & \Sigma   & x_{\Sigma} & \mbox{3-point} \\
 G_5(t,s)& 0 & 1 & 1 & 0 & \Upsilon   \;
 \mbox{and} \; \Gamma & x_{\Upsilon}, x_{\Gamma} & \mbox{4-point} \\ 
 \hline \hline
\end{array}
\]
\caption{\label{tb:contributions} Contributions to the
correlation function: The first column shows how we denote
the contribution, the next four columns give the value
of the corresponding indices. The sixth column lists the
composite field(s) involved, the seventh column how we denote
the scaling dimension of that field. The last column lists
whether it is a three- or four-point function that
contributes.}
\end{table}
A short inspection of the general form of the $n$-points
function given in the appendix B shows that the
contributions have the form (with $y = t/s$)
\BEQ 
{}G_1(t,s) = s^{-x-\frac{1}{2}\tilde{x}_2 + \frac{d}{2}+1}f_1(y)
\;\; , \;\;
{}G_4(t,s) = s^{-x- \frac{1}{2}x_{\Sigma} +\frac{d}{2} + 1} f_4(y)
\EEQ
for the 3-point functions and
\[
\begin{array}{lll}
  G_2(t,s) = s^{-x-x_{\Upsilon} + d+2} f_2(y) & , & 
  G_3(t,s) =  s^{-x-x_{\Gamma} +d + 2} f_3(y) \\  
  G_5(t,s) = s^{-x-\frac{1}{2} x_{\Upsilon} -\frac{1}{2}
  x_{\Gamma} + d+ 2} f_5(y) & & 
\end{array}
\]
for the four-point functions. The scaling functions $f_i(y)$ involve an 
arbitrary functions
$\tilde{\Psi}_i$ which are not fixed by the symmetries 
(see appendix B for details).
As we do not have a free-field theory in this case we can
not make any asumptions about the value of the scaling
dimensions of the composite fields. Therefore we do not
know which terms will be the leading ones in the scaling
regime. However, it turns out that the term $G_1(t,s)$ alone can
reproduce our result correctly. Thus we set the scaling
functions $f_n$=0 with $n=2,\ldots 5$ analogously to the last section. 
We now concentrate on
\BEQ \label{gl:4:G1}
G_1(t,s) = \alpha \rho_0^2 \int \!\D \vec{R} \int \!\D u\: 
\left\langle
\phi(\vec{r},t) \phi(\vec{r},s) \wit{\phi}^2(\vec{R},u)
\right\rangle_0
\EEQ

\subsection{Symmetries of the noiseless theory}

As in the last chapter, we require for the calculation of the  
two- and three-point functions the symmetries of the following non-linear 
`Schr{\"o}dinger equation' obtained from (\ref{gl:sigma02})
\BEQ \label{gl:NLS}
2\mathcal{M}\partial_t \phi(\vec{x},t) = \nabla^2 \phi(\vec{x},t) +
\mathcal{F}(\phi,\wit{\phi})
\EEQ
with a nonlinear potential
\BEQ
\label{gl:nonlinear_term}
\mathcal{F}(\phi,\wit{\phi}) = -g \phi^2(\vec{x},t) \wit{\phi}(\vec{x},t)
\EEQ
While for a constant $g$ the symmetries of this equation are well-known, 
it was pointed out recently that $g$ rather should be considered as a 
dimensionful quantity and hence should transform under local scale-transformations as well \cite{Stoi05}. This requires an extension of 
the generators used so far and we shall give this in appendix~A. 
The computation of the $n$-point functions covariant 
with respect to these new generators is given in
the appendices~B and C. In doing so, we have for technical simplicity 
assumed that to each field
$\vph_i$ there is one associated coupling
constant $g_i$ and only at the end, we let 
\BEQ
g_1 = \ldots =g_n =: g
\EEQ
Therefore, from eq.~(\ref{gl:4:R}) we find for the response function (see (\ref{gl:C:final}))
\BEA
R_0(\vec{r},\vec{r}';t,s) &=& (t-s)^{-\frac{1}{2} (x_1 + x_2)} \left(
\frac{t}{s} \right)^{-\frac{1}{2} (x_1 - x_2)} 
\nonumber \\
& &\times \exp\left( -\frac{\mathcal{M}}{2} \frac{(\vec{r}-\vec{r}')^2}{t-s}
\right) \tilde{\Psi}_2 \left(\frac{t}{s} \cdot \frac{t-s}{g^{1/y}},
\frac{g}{(t-s)^y} \right)
\EEA
with an undetermined scaling function $\tilde{\Psi}_2$. This
form is cleary consistent with our results in table~\ref{tb:results_fun} 
if we identify
\BEQ
x := x_1 = x_2 = a+1 = \frac{d}{2} \;\; , \;\;
\tilde{\Psi}_2 = \mbox{\rm const.}
\EEQ
This holds true for both $\alpha<\alpha_c$ and $\alpha=\alpha_c$. 
In distinction with the bosonic contact process, the symmetries of the
noiseless part $S_0$ do not fix the response function completely but leave a 
certain degree of flexibility in form of the scaling function
$\tilde{\Psi}_2$.  

For the calculation of the correlator we need from eq.~(\ref{gl:4:G1}) 
the following three-point function
\BEA
\left\langle \phi(\vec{r},t) \phi(\vec{r}',s) \tilde{\phi}^2(\vec{R},u)
\right\rangle_0 &=&
(t-s)^{x-\frac{1}{2} \tilde{x}_2} (t-u)^{-\frac{1}{2} \tilde{x}_2}
(s-u)^{-\frac{1}{2} \tilde{x}_2} \nonumber \\ & &
\hspace{-4.0truecm} \times \exp \left( -\frac{\mathcal{M}}{2}
\frac{(\vec{r}-\vec{R})^2}{t-u} - \frac{\mathcal{M}}{2}
\frac{(\vec{r}'-\vec{R})^2}{s - u} \right) \tilde{\Psi}_3 (
u_1,v_1,\beta_1,\beta_2,\beta_3)
\EEA
with
\BEA
u_1 & = & \frac{u}{t}\cdot \frac{ [(s-u)(\vec{r}-\vec{R})
- (t-u)(\vec{r}' - \vec{R})]^2 }{(t-u) (s-u)^2} \\
v_1 & = & \frac{u}{s}\cdot \frac{ [(s-u)(\vec{r}-\vec{R})
- (t-u)(\vec{r}' - \vec{R})]^2 }{(t-u)^2 (s-u)}  \\
\beta_1 & = & \frac{1}{s_2} \cdot \frac{\alpha^{1/y}}{(t-u)^2},
\; \; \beta_2  =  \frac{1}{s_2} \cdot \frac{\alpha^{1/y}}{(s-u)^2},
\;\; \beta_3  =  \alpha^{1/y} s_2 \\
s_2 & = & \frac{1}{t-u} + \frac{1}{u}
\EEA
We choose the following realisation for $\tilde{\Psi}_3$
\BEQ
\label{gl:some_equation}
\tilde{\Psi}_3 (u_1,v_1,\beta_1,\beta_2,\beta_3) = \Xi \left(
\frac{1}{u_1} - \frac{1}{v_1} \right) \left[ -
\frac{(\sqrt{\beta_1}-\sqrt{\beta_2}) \sqrt{\beta_3}
}{\beta_3 - \sqrt{\beta_2 \beta_3} } \right]^{(a-b)}
\EEQ
where the scaling function $\Xi$ was already encountered in eq.~(\ref{gl:3:Xi}) for the bosonic contact process. We now have to distinguish the two different 
cases $\alpha < \alpha_c$ and $\alpha = \alpha_c$. 
For the first case $\alpha<\alpha_c$, we have
$a-b = 0$ so that the last factor in
(\ref{gl:some_equation}) disappears and we simply return to
the expressions already found for the bosonic contact process, in agreement
with the known exact results. However, at the multicritical point
$\alpha=\alpha_c$ we have $a-b \neq 0$ and the last factor becomes
important. We point out that only the  presence or absence of this factor 
distinguishes the cases $\alpha < \alpha_c$ and $\alpha =
\alpha_c$. 

If we substitute the values for $\beta_1,\beta_2$
and $\beta_3$, $\tilde{\Psi}_3$ becomes
\BEQ
\tilde{\Psi}_3 (u_1,v_1,\beta_1,\beta_2,\beta_3) = \Xi \left(
\frac{1}{u_1} - \frac{1}{v_1} \right) 
\left[\frac{\theta (y-1)}{(y-\theta)(1-\theta)}
\right]^{(a-b)}
\EEQ
This factor does not involve $\vec{R}$ so that we obtain in
a similar way as before
\BEA
G_1(t,s) &=& s^{-b} (y-1)^{(b-a)-a-1} \int_0^1 \!\D \theta\,
[(y-\theta)(1-\theta)]^{a-b}\nonumber \\ 
& & \times \phi_1\left( \frac{y+1-2 \theta}{y-1}
\right) \left[ \frac{\theta (y-1)}{(y-\theta) (1-\theta)}
\right]^{a-b} 
\EEA
where we have identified
\BEQ
\tilde{x}_2 = 2(b-a) + d
\EEQ
$G_1(t,s)$ reduces to the expression (\ref{gl:result_int}) if we
choose the same expression for $\phi_1(w)$ as before. We
have thus reproduced all scaling functions correctly.

\section{Conclusions}

The objective of our investigation has been to test further 
the recent proposal of
using the non-trivial dynamical symmetries of a part of the Langevin equation
in order to derive properties of the full stochastic non-equilibrium model. 
To this end, we have compared the known exact results for the two-time
autoresponse and autocorrelation functions in two specific models, see table~\ref{tb:results_fun}, 
with the expressions derived from the
standard field-theoretical actions which are habitually used to describe these
systems. This is achieved through a decomposition of the action into two
parts $S=S_0+S_b$ such that (i) $S_0$ is Schr\"odinger-invariant and the Bargman superselection rules hold for the averages calculated with $S_0$ only and (ii) the remaining terms contained in $S_b$ are such that a perturbative
expansion terminates at a finite order, again due to the Bargman superselection rules. The two models we considered, namely the bosonic variants of the
critical contact and pair-contact processes, satisfy these requirements and are clearly in agreement with
the predictions of local scale-invariance (LSI). In particular, 
our identification eq.~(\ref{gl:phiphit}) of the correct quasi-primary
order-parameter and reponse fields is likely to be useful in more general 
systems. 

Specifically, we have seen the following.
\begin{enumerate}
\item In the bosonic contact process, the symmetries of the noiseless
part $S_0$ of the action is described in terms of the representation of the
Schr\"odinger-group relevant for the {\em free} diffusion equation. 

In consequence, the form of the two-time response function is completely
fixed by LSI and in agreement with the known exact result. The connected
autocorrelator is exactly reducible to certain noiseless three- and four-point
functions. Schr\"odinger-invariance alone cannot determine these but the
remaining free scaling functions can be chosen such that the known exact 
results can be reproduced. 
\item For the bosonic pair-contact process, the
symmetries of the partial action $S_0$ are described in terms of a new
representation pertinent to a {\em non-linear} Schr\"odinger equation. 
This new representation, which we have explicitly constructed, involves a dimensionful coupling constant $g$. 
Therefore even the response function is no longer fully determined. 
As for the autocorrelation function, which again can be exactly reduced to
certain three- and four-point functions calculable from the action $S_0$,
the remaining free scaling functions can be chosen as to fully reproduce
the known exact results. 
\end{enumerate}
The consistency of the predictions of LSI with the exact results of these models
furnishes further evidence in favour of an extension of the well-known
dynamical scaling towards a (hidden) local scale-invariance which
influences the long-time behaviour of slowly relaxing systems. An essential
ingredient were the Bargman superselection rules which at present can only
be derived for a dynamical exponent $z=2$. An extension
of our method to models with $z\ne 2$ would first of
all require a way to generalize the Bargman superselection rules. We hope to
return to this open problem elsewhere. 

\noindent 
{\bf Acknowledgements:} 
F.B. acknowledges the support by the Deutsche
Forschunsgemeinschaft through grant no PL 323/2. 
S.S. was supported by the EU Research Training Network HPRN-CT-2002-00279.
M.H. thanks the Isaac Newton Institute and 
the Universit\"at des Saarlandes 
for warm hospitality, where this work was finished. 
This work was also supported by the Bayerisch-Franz\"osisches
Hochschulzentrum (BFHZ). 

\newpage 

\appsection{A}{Representations of $\mathfrak{age}_1$ and $\mathfrak{sch}_1$
for semi-linear Schr{\"o}dinger equations}
\label{sec:appendixa}

We discuss the Schr{\"o}dinger-invariance of semi-linear
Schr{\"o}dinger equations of the form (\ref{gl:NLS}) and especially with 
non-linearities of the form (\ref{gl:nonlinear_term}). With respect to the
well-known Schr{\"o}dinger-invariance of the linear Schr{\"o}dinger equation, 
the main difference comes from the presence of a dimensionful
coupling constant $g$ of the non-linear term. 

It is enough to consider explicitly the one-dimensional case which 
simplifies the notation. In one spatial dimension,
the Schr\"odinger algebra $\mathfrak{sch}_1$ is spanned by the following
generators
\BEQ \label{gl:A:sch}
\mathfrak{sch}_1=\left\langle X_{-1},X_0,X_1,Y_{-1/2},Y_{1/2},M_0\right\rangle
\EEQ
while its subalgebra $\mathfrak{age}_1$ is spannned by
\BEQ \label{gl:A:age}
\mathfrak{age}_1=\left\langle X_0,X_1,Y_{-1/2},Y_{1/2},M_0\right\rangle
\EEQ
These generators for $g=0$ are listed explicitly in eq.~(\ref{gl:sch1gen})
and the non-vanishing commutators can be written compactly 
\BEA
{[}X_n,X_{n'}] &=& (n-n')X_{n+n'} \nonumber \\ 
{[}X_n,Y_m] &=& (n/2-m)Y_{n+m} \nonumber \\
{[}Y_{\frac{1}{2}},Y_{-\frac{1}{2}}] &= & M_0
\label{gl:A:comm} 
\EEA
where $n,n'\in\{\pm 1,0\}$ and
$m\in\{\pm\frac{1}{2}\}$ (see \cite{Henk02} for generalizations to $d>1$). 

Following the procedure given in \cite{Stoi05}, we now construct new
representations of $\mathfrak{age}_1$ and of $\mathfrak{sch}_1$ which takes into
account a dimensionful coupling $g$ with scaling dimension
$\hat{y}$ as follows. 
\begin{enumerate}
\item The generator of space-translations reads simply 
\BEQ \label{gl:A:Ym}
Y_{-{1\over 2}}=-\partial_r.
\EEQ 
\item The generator of scaling transformations is assumed to take the form
\BEQ \label{gl:A:X0}
X_0=-t\partial_t-{1\over 2}r\partial_r-\hat{y}g\partial_g-{x\over 2}
\EEQ
where $\hat{y}$ is the scaling dimension of the coupling $g$. 
\item For $\mathfrak{sch}_1$ we also keep the usual generator
of time-translations 
\BEQ \label{gl:A:Xm}
X_{-1}=-\partial_t.
\EEQ
\item The remaining generators we write in the most general form adding 
a possible $g$-dependence through yet unknown functions $L,Q,P$.
\BEA
M_0 &=&-{\mathcal M}-L(t,r,g)\partial_g \nonumber \\
Y_{1\over 2} &=&-t\partial_r-\mathcal{M} r-Q(t,r,g)\partial_g
\label{gl:A:Rest} \\
X_1 &=& -t^2\partial_t-tr\partial_r-{\mathcal{M}\over 2}r^2
        -xt-P(t,r,g)\partial_g 
\nonumber 
\EEA
\end{enumerate}  
The representation given by 
eqs.~(\ref{gl:A:Ym},\ref{gl:A:X0},\ref{gl:A:Xm},\ref{gl:A:Rest}) must satisfy 
the commutation relations (\ref{gl:A:comm}) for $\mathfrak{age}_1$ or
$\mathfrak{sch}_1$. From these conditions the undetermined
functions $L,Q$ and $P$ are derived. A straigthforward but slightly longish
calculation along the lines of \cite{Stoi05} shows that for
$\mathfrak{age}_1$, one has
\BEQ \label{eq:resfin}
L=0 \;\; , \;\;  Q=0 \;\; , \;\;
P=p_0(\mathcal{M})\,t^{\hat{y}+1}\,m\left(t/g\right)
\EEQ
Here, $m(v)$ is an arbitrary differentiable function and $p_0(\mathcal{M})$ a
$\mathcal{M}$-dependent constant. We shall
use the shorthand $v = {t^{\hat{y}}}/{g}$ in what follows. 

In consequence, for $\mathfrak{age}_1$ only the generator $X_1$ is
modified with respect to the representation eq.~(\ref{gl:sch1gen}) and this
is described in by the function $m(v)$ and the constant $p_0(\mathcal{M})$.

On the other hand, for $\mathfrak{sch}_1$ the additional condition
$[X_1,X_{-1}]=2X_0$ leads to $p_0=2\hat{y}, m(v)=v^{-1}$.

Hence, the new representations are still given by eq.~(\ref{gl:sch1gen}) with
the only exception of $X_1$ which reads
\BEA
\mathfrak{age}_1 &:~~& X_1 = -t^2 \partial_t - t r\partial_r - p_0(\mathcal{M})
t^{\hat{y}+1} m\left( t^{\hat{y}}/g\right) \partial_g - \frac{\mathcal{M} r^2}{2} - x t
\nonumber \\ 
\mathfrak{sch}_1 &:~~& X_1 = -t^2 \partial_t - t r\partial_r
- 2 \hat{y} t g\partial_g - \frac{\mathcal{M} r^2}{2} - x t 
\label{gl:generatorX1}
\EEA
We require in addition the invariance of linear Schr\"odinger 
equation $(2\mathcal{M}\partial_t-\partial_r^2)\phi=0$ with respect to this 
new representation. In terms of the Schr\"odinger operator $\hat{S}$ this means
\BEQ \label{eq:schinv}
[\hat S,\mathcal{X}]=\lambda \hat S \;\; ; \;\; 
\mbox{\rm where~~~  $\hat{S} := 2M_0X_{-1}-Y_{-1/2}^2$} 
\EEQ
and $\mathcal{X}$ is one of the generators of $\mathfrak{age}_1$
eq.~(\ref{gl:A:age}) or of $\mathfrak{sch}_1$ eq.~(\ref{gl:A:sch}). 
Obviously, $\lambda=0$ if
$\mathcal{X}\in \langle X_{-1},Y_{\pm 1/2},M_0\rangle$ and 
$\lambda=-1$ if $\mathcal{X}=X_0$. 
Finally, for $X_1$ we have 
from the definition of the Schr\"odinger operator $\hat{S}$
\BEA
{}\left[ \hat{S}, X_1 \right] &=& - 4 M_0 X_0 + \left(
Y_{1/2} Y_{-1/2} + Y_{-1/2} Y_{1/2} \right) 
\nonumber \\
&=& -2t \hat{S} + \mathcal{M} \left( 1 - 2x - 4\hat{y} g \partial_g \right)
\EEA
where in the second line the explicit forms 
eqs.~(\ref{gl:A:Ym},\ref{gl:A:X0},\ref{gl:A:Xm},\ref{gl:A:Rest})
were used. This also holds for all those representations of $\mathfrak{age}_1$
for which there exists an operator 
$X_{-1}\not\in\mathfrak{age}_1$ such that $[X_1,X_{-1}]=2X_0$ 
and we shall restrict our attention to those in what follows. 
On the other hand, the direct calculation of the same commutator with the
explicit form (\ref{gl:generatorX1}) gives for $\mathfrak{age}_1$
\BEQ
\hspace{-2truecm} 
{}\left[ \hat{S}, X_1 \right] = -2t\hat{S} + \mathcal{M}\left( 1 - 2x\right)
-2\mathcal{M}p_0(\mathcal{M}) t^{\hat{y}} \left[ 
(\hat{y}+1) m(v) + \hat{y} v m'(v))\right] \partial_g
\EEQ 
Besides $\lambda=-2t$, the consistency between these two implies for 
$m(v)$ the equation
\BEQ
v \left( (\hat{y}+1) m(v) + \hat{y} v \frac{\D m(v)}{\D v} \right) = 
\frac{2\hat{y}}{p_0} 
\EEQ
with the general solution
\BEQ
m(v) = \frac{2 \hat{y}}{p_0} \frac{1}{v} + \frac{m_0}{p_0}\,
v^{-1-1/\hat{y}}
\EEQ
where $m_0=m_0(\mathcal{M})$ is an arbitrary constant. 
The larger algebra $\mathfrak{sch}_1$ is recovered from this if we set
$p_0=2\hat{y}$ and $m_0=0$. 
Hence the final form for the
generator $X_1$ in the special class of representations of the algebra $\mathfrak{age}_1$ defined above is
\BEQ \label{gl:A:X1final}
X_1 = - t^2 \partial_t - t r\partial_t - 2 \hat{y} t g\partial_g -  m_0
g^{1+1/\hat{y}}\partial_g -\frac{\mathcal{M} r^2}{2} - x t
\EEQ
Summarizing, this class of representations of $\mathfrak{age}_1$ we 
constructed is characterized by the triplet $(x,\mathcal{M},m_0)$, whereas for 
$\mathfrak{sch}_1$, the same triplet is $(x,\mathcal{M},0)$. 

Finally, to make $X_1$ a dynamical symmetry on the solutions 
$\Phi=\Phi_g(t,r)$ of the Schr\"odinger equation $\hat{S} \Phi_g=0$ we must
impose the auxiliary condition $(1-2x-4\hat{y}g\partial_g)\Phi_g=0$ 
which leads to
\BEQ \label{eq:partem}
\Phi_g(t,r)=g^{(1-2x)/(4\hat{y})}\Phi(t,r) 
\EEQ
In particular, we see that if $x=1/2$, we have  a representation of 
$\mathfrak{age}_1$ without any further auxiliary condition. 

We now look for those semi-linear Schr\"odinger equations of the
form $\hat{S}\Phi = F(t,r,g,\Phi,\Phi^*)$ for which the representations
of $\mathfrak{age}_1$ or $\mathfrak{sch}_1$ as given by
eqs.~(\ref{gl:A:Ym},\ref{gl:A:X0},\ref{gl:A:Xm},\ref{gl:A:Rest}) and
with $X_1$ as in (\ref{gl:A:X1final}) act as a dynamical symmetry. 
The non-linear potential $F$ is known to satisfy certain differential
equations which can be found using standard methods, see 
\cite{Boye76},\cite[eq. (2.8)]{Stoi05}. In our case these equations read
\BEA
X_{-1}          &:~& \partial_t F = 0 \label{eq:tm} \\
Y_{-{1\over 2}} &:~& \partial_r F = 0 \label{eq:sp} \\
M_0             &:~& (\Phi\partial_{\Phi}-\Phi^*\partial_{\Phi^*}-1)F = 0 
                    \label {eq:mas}\\
Y_{1\over 2}    &:~& \left[ t\partial_rF - 
                    \mathcal{M}r(\Phi\partial_{\Phi}-\Phi^*\partial_{\Phi^*}
                    -1)\right]F= 0\label{eq:gl} \\
X_0             &:~& \left[t\partial_t +{1\over
2}r\partial_r+\hat{y} g\partial_g +1 
                   -{x\over 2}(\Phi \partial_{\Phi}+\Phi^*\partial_{\Phi^*}-1)
                   \right] F = 0 
\label{eq:dl} \\
X_1             &:~& \Big[
t^2\partial_t+tr\partial_r+2t(\hat{y}g\partial_g +1)
                    +m_0 g^{1+1/\hat{y}}\partial_g
		    \nonumber \\ & & 
                    -\frac{\mathcal{M}r^2}{2}
		    (\Phi\partial_{\Phi}-\Phi^*\partial_{\Phi^*}-1)
		  -xt(\Phi\partial_{\Phi}+\Phi^*\partial_{\Phi^*}-1)
		  \Big]F = 0\label{eq:cf} 
\EEA

We first solve these for $\mathfrak{sch}_1$. From the conditions
eqs.~(\ref{eq:tm},\ref{eq:sp},\ref{eq:mas},\ref{eq:gl},\ref{eq:dl}) we easily 
find 
\BEQ \label{gl:A:Fsch1}
F = \Phi \left( \Phi \Phi^*\right)^{1/x} f\left( g^x
\left( \Phi \Phi^*\right)^{\hat{y}}\right)
\EEQ
where $f$ is an arbitrary differentiable function. Two comments are in order:
\begin{enumerate}
\item For a dimensionless coupling $g$, that is $\hat{y}=0$,
we have $x=1/2$. Then
the scaling function reduces to a $g$-dependent constant and we 
recover the standard form for the non-linear
potential $F$ as quoted ubiquitously in the mathematical literature, 
see e.g. \cite{Fush93}.
\item Taking into account the generator $X_1$ from eq.~(\ref{eq:cf}) 
as well does not change the
result. Hence in this case translation-, dilatation- and Galilei-invariance
are indeed sufficient for the special Schr\"odinger-invariance generated by
$X_1$, see also \cite{Henk03}. We point out that traditionnally an analogous
assertion holds for conformal field-theory, see e.g. \cite{Card96}, but 
counterexamples are known where in local theories scale- and translation-invariance are {\em not} sufficient for conformal invariance
\cite{Polc88,Riva05}.
\end{enumerate} 

Second, we now consider the representation of $\mathfrak{age}_1$ where $X_1$ is given by (\ref{gl:A:X1final}). We have the conditions
eqs.~(\ref{eq:sp},\ref{eq:mas},\ref{eq:gl},\ref{eq:dl},\ref{eq:cf}). We write
$F=\Phi \mathcal{F}(\omega,t,g)$ with $\omega=\Phi\Phi^*$ and the remaining
equations coming from $X_0$ and $X_1$ are
\BEA
\left( t\partial_t + \hat{y} g \partial_g - x u\partial_u +1 \right)\mathcal{F} &=& 0
\nonumber \\
\left( t^2 \partial_t + m_0 g^{1+1/\hat{y}}\partial_g \right) \mathcal{F} &=& 0
\EEA
with the final result
\BEQ
F = \Phi \left( \Phi \Phi^*\right)^{1/x} f\left( \left( \Phi
\Phi^*\right)^{\hat{y}}
\left[g^{-1/\hat{y}}-\frac{m_0}{\hat{y}\,t}\right]^{-x\hat{y}}\right)
\EEQ
and where $f$ is the {\em same} scaling function as encountered before for
$\mathfrak{sch}_1$. Finally, the result for the general representations of 
$\mathfrak{age}_1$ which depend on an arbitrary function $m(v)$ are not
particularly inspiring and will not be detailed here. We observe
\begin{enumerate}
\item For $m_0=0$, this result is identical to the one found for 
$\mathfrak{sch}_1$. 
\item Even for $m_0\ne 0$, the form of the non-linear potential reduces in 
the long-time limit $t\to\infty$ to the one found in eq.~(\ref{gl:A:Fsch1}) 
for the larger algebra $\mathfrak{sch}_1$. 
\end{enumerate}
We can summarize the main results of this appendix as follows.\\

\noindent {\bf Proposition.} {\it Consider the following generators}
\BEA
\label{generators}
M_0 &=& - \mathcal{M} \;\; , \;\; Y_{-1/2} \:=\: -\partial_r \;\; , \;\;
Y_{1/2} \:=\: - t\partial_r - \mathcal{M} r 
\;\; , \;\; X_{-1} \:=\: -\partial_{t}
\nonumber \\
X_0 &=& -t\partial_t - \frac{1}{2} r\partial_r - \hat{y} 
g\partial_g - \frac{x}{2}
\\
X_1 &=& - t^2 \partial_t - t r\partial_t - 2\hat{y}t
g\partial_g -  m_0 g^{1+1/\hat{y}}\partial_g -\frac{\mathcal{M} r^2}{2} - x t
\nonumber
\EEA
{\it where $x,\mathcal{M},m_0$ are parameters. Define the
Schr\"odinger operator $\hat{S} := 2M_0 X_{-1} - Y_{-1/2}^2$. Then:\\
(i) the generators $\langle X_{0,1},Y_{\pm 1/2},M_0\rangle$ form a representation of the 
Lie algebra $\mathfrak{age}_1$. If furthermore $m_0=0$, then 
$\langle X_{0,\pm 1},Y_{\pm 1/2},M_0\rangle$ is a representation of the
Lie algebra $\mathfrak{sch}_1$. \\
(ii) These representations are dynamical symmetries of the Schr\"odinger equation $\hat{S}\Phi=0$, under the auxiliary condition
$(1-2x-4\hat{y} g\partial_g)\Phi=0$. \\
(iii) For the Schr\"odinger-algebra $\mathfrak{sch}_1$ and also in the
asymptotic limit $t\to\infty$ for the ageing algebra $\mathfrak{age}_1$, the
semi-linear Schr\"odinger equation invariant under these representations has the form}
\BEQ
\hat{S} \Phi = \Phi \left( \Phi \Phi^*\right)^{1/x} f\left( g^x \left(
\Phi \Phi^*\right)^{\hat{y}}\right)
\EEQ
{\it where $f$ is an arbitrary differentiable function.}

This general form includes our potential
(\ref{gl:nonlinear_term}) since the scaling dimension
$\hat{y}$ is a
remaining free parameter in our considerations

\appsection{B}{The $n$-point function}
\label{appb}
In this appendix we use the generators (\ref{generators}) from  appendix~A to 
find the most general form of the $n$-point functions compatible
with the symmetries for $n \geq 3$. 
We shall do this for the case $m_0 = 0$ only, as this will be enough to reproduce the exact results of table~\ref{tb:results_fun}. The case $n=2$
needs a special treatment and is presented in appendix~C. 

We restrict ourselves
to the case $d=1$ for simplicity, but the generalisation 
to arbitrary dimension will be obvious. First we introduce
some notation. We fix an arbitrary index $k$ and 
define the shifted coordinates
\BEQ
\tilde{\vec{r}}_b := \vec{r}_b - \vec{r}_k, \quad 
\tilde{t}_b := t_b - t_k \quad \mbox{for} \quad j \neq k
\quad \mbox{and} \quad \tilde{t}_k := t_k
\EEQ
In the sequel, we will adopt the following convention: The
index $a$ always runs from $1$ to $n$, the index $b$ runs
from $1$ to $n$ but skips $k$. 
The prime on a sum means that the index $k$ is left out, viz. 
\BEQ
{\sum_{i=1}^n}' A_i := \sum_{\stackrel{i=1}{i \neq k}}^n A_i
\EEQ

We denote the $n$-point function by
\BEQ
\label{gl:appb:definition}
F(\{\vec{r}_a\},\{t_a\},\{g_a\}) := \langle \vph_1(\vec{r}_1,t_1) \ldots
\vph_n(r_n,t_n) \rangle 
\EEQ
where we assume one coupling constant for each field.
This quantity has to satisfy the following four linear 
partial differential equations
\BEA
\label{gl:appb:equations}
\left(\sum_{i=1}^n X_k^{(i)}\right)
F(\{\vec{r}_a\},\{t_a\},\{t_a\}) &=& 0 \;\; ; \;\; k \in \{0,1\} \\ 
\left(\sum_{i=1}^n Y_m^{(i)}\right)
F(\{\vec{r}_a\},\{t_a\},\{t_a\}) &=& 0 
\;\; ; \;\; m \in \left\{-\frac{1}{2},\frac{1}{2} \right\}
\EEA
To solve these equations, we use the method of
characteristics \cite{Kamk59}.  
We solve (\ref{gl:appb:equations}) first for spatial
translation invariance with the result
\BEQ
F(\{\vec{r}_a\},\{t_a\},\{g_a\}) =
\tilde{F}(\{\tilde{\vec{r}}_b\},\{t_a\},\{g_a\}) 
\EEQ
with a new function $\tilde{F}$ with $3 n -1$ arguments.
In order to solve for $X_0$ we set
\BEQ
x = \frac{1}{2}\sum_{i=1}^n x_i
\EEQ
and make the ansatz
\BEQ
\label{gl:appb:ansatz1}
\tilde{F}(\{\tilde{\vec{r}}_b\},\{t_a\},\{g_a\}) 
= \prod_{i<j} (t_i - t_j)^{-\rho_{ij}}
G(\{\tilde{\vec{r}}_b\},\{\tilde{t}_a\},\{g_a\})
\EEQ
where the parameters $\rho_{ij}$ and the function $G$
remain to be determined. We
also change to the new independent temporal variables
$\tilde{t}_a$. Then one finds after a short calculation
\BEQ
\label{gl:appb:x02}
\left( \sum_{i=1}^n \tilde{t}_i \partial_{\tilde{t}_i} +
\frac{1}{2} {\sum_{i=1}^n}' \tilde{\vec{r}}_i
\partial_{\tilde{\vec{r}}_i} + \sum_{i=1}^n \hat{y}_i g_i \partial_{g_i}
\right) G(\{\tilde{\vec{r}}_b\},\{\tilde{t}_a\},\{g_a\}) = 0
\EEQ
together with the condition
\BEQ
\label{gl:appb:cond1}
x = \sum_{i < j} \rho_{ij}.
\EEQ
Before proceeding to solve this equation, we turn to the
generators $Y_{{1}/{2}}$. We find for the function
$G(\{\tilde{\vec{r}}_b\},\{\tilde{t}_a \},\{g_a\} )$
\BEQ
\label{gl:appb:y12}
\left( {\sum_{i=1}^n}' \tilde{t}_i \partial_{\tilde{\vec{r}}_i} +
{\sum_{i=1}^n}'\mathcal{M}_i \tilde{\vec{r}}_i + \vec{r}_k \left(\sum_{i=1}^n
\mathcal{M}_i \right)\right)  G(\{\tilde{\vec{r}}_b\},\{\tilde{t}_a \},\{g_a\} )
= 0.
\EEQ
Since $G(\{\tilde{\vec{r}}_b\},\{\tilde{t}_a \},\{g_a\})$ does 
not depend on $\vec{r}_k$, we recover the Bargman superselection rule
\BEQ
\label{gl:appb:bargmann}
\sum_{i=1}^n \mathcal{M}_i = 0
\EEQ
as expected. 
For $G(\{\tilde{\vec{r}}_b\},\{\tilde{t}_a \},\{g_a\})$ we make another ansatz:
\BEQ
\label{gl:appb:ansatz2}
G(\{\tilde{\vec{r}}_b \},\{\tilde{t}_a \},\{g_a \}) =
\exp \left(-{\sum_{i=1}^n}'\frac{\mathcal{M}_i}{2} 
\frac{\tilde{\vec{r}}_i^2}{\tilde{t}_i} \right) 
H(\{\tilde{\vec{r}}_b \},\{\tilde{t}_a\},\{g_a\})
\EEQ
where the function $H(\{\tilde{\vec{r}}_b\},\{\tilde{t}_a
\},\{g_a\})$ remains to be determined. With
(\ref{gl:appb:bargmann}) and (\ref{gl:appb:ansatz2}),
equation (\ref{gl:appb:y12}) reduces to
\BEQ
\label{gl:appb:h1}
\left( {\sum_{i=1}^n}' \tilde{t}_i
\partial_{\tilde{\vec{r}}_i} \right) H(\{\tilde{\vec{r}}_b\},
\{\tilde{t}_a\},\{g_a\}) = 0.
\EEQ
We retake (\ref{gl:appb:x02}) and introduce the ansatz
(\ref{gl:appb:ansatz2}). This yields 
\BEQ
\label{gl:appb:h2}
\left( \sum_{i=1}^n \tilde{t}_i \partial_{\tilde{t}_i} +
\frac{1}{2} {\sum_{i=1}^n}' \tilde{\vec{r}}_i
\partial_{\tilde{\vec{r}}_i} + \sum_{i=1}^n \hat{y}_i g_i
\partial_{g_i} \right) H(\{\tilde{\vec{r}}_b \},\{
\tilde{t}_a \},\{ g_a \}) = 0
\EEQ
The last equation for $H(\{\tilde{\vec{r}}_b\},\{\tilde{t}_a
\},\{g_a\})$ we obtain from $X_1$. Using
the ansatz (\ref{gl:appb:ansatz1}) yields an equation for $G$
\BEA
\label{gl:appb:x12}
& &\hspace{-2.5cm}
\left(\sum_{i=1}^n \tilde{t}_i^2 \partial_{\tilde{t}_i} +
{\sum_{i=1}^n}' \tilde{t}_i
\tilde{\vec{r}}_i \partial_{\tilde{\vec{r}}_i}+ \frac{1}{2}
{\sum_{i=1}^n}' \mathcal{M}_i \tilde{\vec{r}}_i^2 +
2 {\sum_{i=1}^n}' \hat{y}_i \tilde{t}_i g_i \partial_{g_i}
+\vec{r}_k \left( {\sum_{i=1}^n}' \tilde{t}_i
\partial_{\tilde{\vec{r}}_i} + {\sum_{i=1}^n}'
\tilde{\vec{r}_i} \mathcal{M}_i \right) \right. \nonumber \\
& & \hspace{-1.5cm} \left. +2 \tilde{t}_k \left({\sum_{i=1}^n}' \tilde{t}_i
\partial_{\tilde{t}_i} + \frac{1}{2} {\sum_{i=1}^n}'
\tilde{\vec{r}}_i \partial_{\vec{r}_i} + {\sum_{i=1}^n}'
\hat{y}_i
g_i \partial_{g_i} \right)  
\right) G(\{\tilde{\vec{r}}_b \},\{ \tilde{t}_a \},\{ g_a
\})= 0.
\EEA
Together with the condition
\BEQ
\sum_{i < j} (t_i + t_j) = \sum_{i=1}^n t_i x_i
\EEQ
which is satisfied if we choose the parameters $\rho_{ij}$
such that
\BEA
\label{gl:appb:system}
x_1 &=& \rho_{12}+ \rho_{13}+\rho_{14}+\rho_{15}+ \ldots+\rho_{1n} \nonumber \\
x_2 &=& \rho_{12} + \rho_{23} + \rho_{24} + \rho_{25}+
\ldots +\rho_{2n} \nonumber \\
x_3 &=& \rho_{13} + \rho_{23} + \rho_{34} + \rho_{35}+
\ldots + \rho_{3n} \\
\vdots &  & \vdots \qquad \qquad \qquad \vdots \qquad \qquad
\vdots\nonumber  \\
x_n &=& \rho_{1n} + \rho_{2n} + \rho_{3n} + \rho_{4n}+ \ldots + \rho_{n-1\,n}
\nonumber
\EEA
Here a few remarks are in order. The above system is
compatible with (\ref{gl:appb:cond1}), as can be see be adding all equations.
Also, this system is always solvable for $n \geq 3$, as for $n
\geq 4$, it is underdetermined and for $n = 3$ the
correponding determinant does not vanish \footnote{This system is
not solvable for $n=2$ when $x_1\ne x_2$. 
This case is considered in appendix~C.}. Lastly, we often
have the case $x_1 = x_2 =: x$ and $x_3 = \ldots = x_n =:
\tilde{x}$. In this case, we can set
\BEQ
\label{special_case}
\begin{array}{cccr}
\rho_{12} = x - \frac{n-2}{2} \tilde{x} ~;~~ & 
\rho_{2 i}= \frac{1}{2} \tilde{x}, & \rho_{1 i} = \frac{1}{2} \tilde{x} & 
\mbox{\rm for $i = 3, \ldots , n$}
\end{array}
\EEQ
and $\rho_{ij} = 0$ for all the remaining $\rho_{ij}$.
We still have to rewrite equation (\ref{gl:appb:x12}) in terms of 
the variables $\{ \tilde{\vec{r}}_b \}$ and $\{ \tilde{t}_a
\}$. Here we take equations (\ref{gl:appb:x02}) and
(\ref{gl:appb:y12}) and the ansatz (\ref{gl:appb:ansatz2}) 
into account and get for $H(\{\tilde{\vec{r}}_b \}
,\{ \tilde{t}_a \},\{ g_a \})$
\BEQ
\label{gl:appb:h3}
\hspace{-2.0truecm}
\left( {\sum_{i=1}^n}' \tilde{t}_i^2
\partial_{\tilde{t}_i} - \tilde{t}_k^2 \partial_{t_k} +
{\sum_{i=1}^n}' \tilde{t}_i \tilde{\vec{r}}_i
\partial_{\tilde{\vec{r}}_i} + 2 {\sum_{i=1}^n}'
\hat{y}_i \tilde{t}_i g_i \partial_{g_i} \right) H(\{\tilde{\vec{r}}_b \}
,\{ \tilde{t}_a \},\{ g_a \}) =  0.
\EEQ
We thus have to solve the homogenous equations
(\ref{gl:appb:h1}),(\ref{gl:appb:h2}) and
(\ref{gl:appb:h3}). This will eliminate three more variables
and yields
\BEA
F(\{\vec{r}_a\},\{t_a\},\{g_a\}) &=& \prod_{i < j} (t_i -
t_j)^{-\rho_{ij}} \exp \left( -\frac{1}{2}{\sum_{i=1}^n}'
\mathcal{M}_i \frac{(\vec{r}_i - \vec{r}_k)^2}{t_i - t_k} \right)
\nonumber \\
& & \times \tilde{\Psi}_n \left( \{u_c\},\{v_c\},\{\beta_a\} \right)
\EEA 
with an arbitrary function $\tilde{\Psi}_n$, which depends on $3 n -
4$ variables. Here the index $c$ runs from $1$ to $n$ but
skips $k$ and another arbitrarily fixed index $r \neq k$,
and the expressions $u_c,v_c$ and $\beta_a$ are
given by
\BEA
u_c &=& \frac{t_k \left( (\vec{r}_c - \vec{r}_k) (t_r - t_k)
- (\vec{r}_r - \vec{r}_k)
(t_c - t_k) \right)^2}{(t_c - t_k) (t_r - t_k)^2 t_c}
\nonumber \\
v_c &=& \frac{t_k \left( (\vec{r}_c - \vec{r}_k) (t_r - t_k)
- (\vec{r}_r - \vec{r}_k)
(t_c - t_k) \right)^2}{(t_r - t_k) (t_c - t_k)^2 t_r}
\nonumber \\
\beta_k &=& {g_k}^{(1/\hat{y}_k)} \left( \frac{t_r}{(t_r - t_k) t_k}
\right) ,\qquad \beta_b = {g_b}^{(1/\hat{y}_b)} \left( \frac{t_k (t_b -
t_k)^2}{(t_r - t_k) t_r} \right) 
\EEA
We remind the reader of our convention that the index $c$
runs from $1$ to $n$ skipping $r$ and $k$ and that the index
$b$ runs from $1$ to $n$ skipping only $k$.

In higher dimensions rotational invariance has to be
satisfied as well and then the generalization to arbitrary $d$ is
straightforward.

If we consider instead the algebra $\mathfrak{age}_1$ with dimensionless
couplings $g_i$ we merely have to make the replacement
\BEQ
\label{gl:appb:result2}
\tilde{\Psi}_n \left( \{u_c\},\{v_c\},\{\beta_a\} \right)
\longrightarrow 
\Psi_n \left( \{u_c\},\{v_c\}\right)
\EEQ
where $\Psi_n$ is also an arbitrary function such that 
only the dependence on $\{ \beta_a \}$ drops out.

Finally, we explicitely list the three- and four-point
functions in the form in which they are needed in the main text. 
The three-point function with fixed indices $r = 2$ and $k = 3$ and the
special situation (\ref{special_case}) assumed reads
\BEA
\label{app_threepoint}
F(\{\vec{r}_a\},\{t_a\},\{g_a\}) &=&
(t_1-t_2)^{-(x-\frac{1}{2} \tilde{x})} (t_1 -
t_3)^{-\frac{1}{2} \tilde{x}} (t_2-t_3)^{-\frac{1}{2}
\tilde{x}} \\
& & \hspace{-1.0cm} \times \exp \left( -\frac{1}{2}{\sum_{i=1}^2}
\mathcal{M}_i \frac{(\vec{r}_i - \vec{r}_3)^2}{t_i - t_3} \right)
\tilde{\Psi}_n \left( \{u_c\},\{v_c\},\{\beta_a\} \right)
\nonumber
\EEA 
with 
\BEA
u_1 & = & \frac{t_3}{t_1} \frac{[(\vec{r}_1 - \vec{r}_3)(t_2
- t_3) - (\vec{r}_2 -\vec{r}_3)(t_1 - t_3)]^2}{(t_1
-t_3)(t_2 - t_3)^2} \nonumber \\
v_1 & = &\frac{t_3}{t_2} \frac{[(\vec{r}_1 - \vec{r}_3)(t_2
- t_3) - (\vec{r}_2 -\vec{r}_3)(t_1 - t_3)]^2}{(t_1
-t_3)^2 (t_2 - t_3)} \\
\beta_1 &=& {g_1}^{1/\hat{y}_1} \frac{t_3 (t_1 - t_3)^2}{(t_2 - t_3)
t_2}, \qquad \beta_2 = {g_2}^{1/\hat{y}_2} \frac{t_3 (t_2 -
t_3)}{t_2} \nonumber \\
\beta_3 &=& {g_3}^{1/\hat{y}_3} \frac{t_2}{t_3 (t_2 - t_3)}
\nonumber
\EEA
The four point function with $r=1$ and $k=2$ in the special
situation (\ref{special_case}) reads
\BEA
\hspace{-2.0cm}
F(\{\vec{r}_a\},\{t_a\},\{g_a\}) &=&
(t_1-t_2)^{-(x- \tilde{x})} (t_1 -
t_3)^{-\frac{1}{2} \tilde{x}} (t_1 -
t_4)^{-\frac{1}{2}\tilde{x}} (t_2-t_3)^{-\frac{1}{2} \tilde{x}}
\nonumber \\ 
& & \hspace{-2.0cm} \times (t_2 - t_4)^{-\frac{1}{2}\tilde{x}}  
\exp \left( -\frac{1}{2}{\sum_{i=1}^3}
\mathcal{M}_i \frac{(\vec{r}_i - \vec{r}_4)^2}{t_i - t_4} \right)
\tilde{\Psi}_n \left( \{u_c\},\{v_c\},\{\beta_a\} \right)
\nonumber
\EEA 
with 
\BEA
\label{app_fourpoint}
u_3 & = & \frac{t_2}{t_3} \frac{[(\vec{r}_3 - \vec{r}_2)(t_1
- t_2) - (\vec{r}_1 -\vec{r}_2)(t_3 - t_4)]^2}{(t_3
-t_2)(t_1 - t_2)^2} \nonumber \\
u_4 & = & \frac{t_2}{t_4} \frac{[(\vec{r}_4 - \vec{r}_2)(t_1
- t_2) - (\vec{r}_1 -\vec{r}_2)(t_4 - t_2)]^2}{(t_4
-t_2)(t_1 - t_2)^2} \nonumber \\
v_3 & = &\frac{t_2}{t_1} \frac{[(\vec{r}_3 - \vec{r}_2)(t_1
- t_2) - (\vec{r}_1 -\vec{r}_2)(t_3 - t_2)]^2}{(t_1
-t_2) (t_3 - t_2)^2} \\
v_4 & = &\frac{t_2}{t_1} \frac{[(\vec{r}_4 - \vec{r}_2)(t_1
- t_2) - (\vec{r}_1 -\vec{r}_2)(t_4 - t_2)]^2}{(t_1
-t_2) (t_4 - t_2)^2} \nonumber \\
\beta_1 &=& {g_1}^{1/\hat{y}_1} \frac{t_2 (t_1 - t_2)}{t_1}, 
\qquad \beta_2 = {g_2}^{1/\hat{y}_2} \frac{t_1}{(t_1-t_2) t_2}
\nonumber \\
\beta_3 &=& {g_3}^{1/\hat{y}_3} \frac{t_2(t_3 - t_2)^2}{t_1 (t_1 -
t_2)},
\qquad \beta_4 = {g_4}^{1/\hat{y}_4} \frac{t_2 (t_4-t_2)^2}{(t_1 -
t_2) t_1} \nonumber
\EEA

\appsection{C}{The two-point function}

In this appendix we calculate  the two-point
function, which was not included in the treatment of
appendix~B. Again, we only treat the case $m_0 = 0$.
Apart from the generator $X_1$, the calculations are similar to those done in
appendix~B, so we only give the essential steps.
First we define
\BEQ
\tau := t_1-t_2 \;\; , \;\; \vec{r} := \vec{r}_1 - \vec{r}_2
\EEQ
and then we proceed as follows. We solve for
$M_0,Y_{-{1}/{2}}, Y_{{1}/{2}}, X_0$ in exactly the same
way as before with the result
\BEQ
\hspace{-2truecm}
F(\vec{r}_1,\vec{r}_2,t_1,t_2,g_1,g_2) = 
\left\langle \vph_1(\vec{r}_1,t_1,g_1) \vph_2(\vec{r}_2,t_2,g_2)
\right\rangle_0 = \tau^{-x} G(\vec{r},\tau,t_2,g_1,g_2)
\EEQ
where $x = \frac{1}{2} \left(x_1 + x_2 \right)$ and 
$G(\vec{r},\tau,t_2,g_1,g_2)$ satisfies the equations
\BEA
\label{gl:appc:g1}
\left(\tau \partial_\tau + t_2 \partial_{t_2} + \frac{1}{2}
\vec{r} \partial_{\vec{r}} + y_1 g_1 \partial_{g_1} + y_2 g_2
\partial_{g_2} \right) G(\vec{r},\tau,t_2,g_1,g_2) &=& 0 \\
\label{gl:appc:g2}
(\tau \partial_{\vec{r}} + \vec{r} \mathcal{M}_1) G(\vec{r},\tau,t_2,g_1,g_2)
& = & 0  
\EEA
and the Bargman superselection rule 
\BEQ
\mathcal{M}_1 + \mathcal{M}_2 = 0
\EEQ
holds true. Now
(\ref{gl:appc:g1}) is solved by
\BEQ
G(\vec{r},\tau,t_2,g_1,g_2) = \tilde{G}(u_1,u_2,v_1,v_2)
\EEQ
where we have defined
\BEQ
u_1 := \frac{\vec{r}^2}{\tau}, \quad u_2 := \frac{\vec{r}^2}{t_2}, \quad
v_1 := \frac{g_1^{1/\hat{y}_1}}{\tau}, \quad v_2 :=
\frac{g_2^{1/\hat{y}_2}}{\tau}
\EEQ
and rewriting (\ref{gl:appc:g2}) in terms of the new variables
yields
\BEQ
\left(u_1 \partial_{u_1} + u_2 \partial_{u_2} + \frac{1}{2} u_1
\mathcal{M}_1 \right) \tilde{G}(u_1,u_2,v_1,v_2)= 0
\EEQ
which is solved by
\BEQ
\tilde{G}(u_1,u_2,v_1,v_2) = \exp \left( - \frac{1}{2}u_1 \mathcal{M}_1
\right) H(w,v_1,v_2) \;\; , \;\; w:= \frac{u_2}{u_1}
\EEQ
The function $H(w,v_1,v_2)$ is found through the generator $X_1$. 
Using again the invariance under $Y_{{1}/{2}}$  and $X_0$, we readily 
obtain in terms of $v_1,v_2$ and $w$
\BEQ
\left((w+1) \partial_w + v_1 \partial_{v_1} - v_2 \partial_{v_2}
+ \frac{1}{2}(x_1 - x_2) \right) H(w,v_1,v_2) = 0.
\EEQ
The most general solution of this equation is
\BEQ
H = (w + 1)^{-\frac{1}{2}(x_1-x_2)}
\tilde{\Psi}_2 \left( \frac{(w+1)}{v_1},v_1 v_2 \right)
\EEQ
where the function $\tilde{\Psi}_2$ remains arbitrary. 
Substituting back the values for $v_1,v_2$ and $w$ our final result is
\BEA
\lefteqn{ 
F(\vec{r}_1,t_1,r_2,t_2) = \delta_{\mathcal{M}_1 + \mathcal{M}_2,0}\,
(t_1 - t_2)^{-\frac{1}{2}(x_1 + x_2)}
\left( \frac{t_1}{t_2} \right)^{-\frac{1}{2}(x_1-x_2)}
}
\nonumber  \\ & & 
\times \exp \left( -\frac{\mathcal{M}_1}{2} 
\frac{(\vec{r}_1-\vec{r}_2)^2}{t_1 - t_2}\right)  
\tilde{\Psi}_2 \left(
\left(\frac{t_1}{t_2}\right)^{\hat{y}_1}\frac{(t_1-t_2)^{\hat{y}_1}}{g_1}\, , \,
\frac{g_1 g_2}{(t_1-t_2)^{\hat{y}_1 + \hat{y}_2}} \right).
\label{gl:C:final}
\EEA
For applications to semi-linear equations, one now sets $g:=g_1=g_2$ with
a scaling dimension $\hat{y}:=\hat{y}_1=\hat{y}_2$. In the
limit $\hat{y} \to 0$, the function
$\tilde{\Psi}_2$ reduces to a $g$-dependent normalization constant 
and we recover the standard result \cite{Henk94}. 

In many applications, one expects the scaling functions to be universal,
up to normalization. On the other hand, the coupling $g$ should be 
a non-universal quantity so that a universal scaling function cannot
contain $g$ in its arguments. This leads to
$\tilde{\Psi}_2 = \tilde{\Psi}_2( (t_1/t_2)^{\hat{y}})$
and we point out that such a scaling form would be compatible (one still
has $z=2$, however) with what
is found from the field-theoretical renormalization group and numerical
simulations in non-equilibrium critical dynamics \cite{Cala05,Plei05}. 
An extension to different values of $z$ would as a first step require
the generalization of the Bargman superselection rules. We hope to come
back elsewhere to this open problem. 
\newpage 

  
\end{document}